%% file: paper.tex
\documentclass[acmsmall]{acmart}

\usepackage{multirow}
\usepackage{adjustbox}
\usepackage{framed}
\usepackage{Sweave}
\usepackage{balance} 
\usepackage{listings}
\usepackage{cprotect}
\usepackage{hyperref}

\definecolor{dkgreen}{rgb}{0,0.6,0}
\definecolor{gray}{rgb}{0.5,0.5,0.5}
\definecolor{mauve}{rgb}{0.58,0,0.82}
\graphicspath{{./figures/}}

\AtBeginDocument{%
  \providecommand\BibTeX{{%
    \normalfont B\kern-0.5em{\scshape i\kern-0.25em b}\kern-0.8em\TeX}}}

\setcopyright{acmcopyright}
\copyrightyear{2021}
\acmYear{2021}
\acmDOI{10.1145/9999999.9999999}

\acmJournal{TOSEM}
\acmVolume{99}
\acmNumber{9}
\acmArticle{999}
\acmMonth{9}

\begin{document}

\title{Test cases as a measurement instrument in experimentation}

\author{O. Dieste}
\email{odieste@fi.upm.es}
\orcid{0000-0002-3060-7853}
\affiliation{%
  \institution{Universidad Polit\'ecnica de Madrid}
  \city{Boadilla del Monte}
  \postcode{28660}
  \country{Spain}
}

\author{F. Uyaguari}
\email{fuyaguar@etapa.net.ec}
\orcid{0000-0001-7060-1002 }
\affiliation{%
  \institution{ETAPA Telecommunications Company}
  \city{Cuenca}
  \postcode{10204}
  \country{Ecuador}
}

\author{S. Vegas}
\email{svegas@fi.upm.es}
\orcid{0000-0001-8535-9386}
\affiliation{%
  \institution{Universidad Polit\'ecnica de Madrid}
  \city{Boadilla del Monte}
  \postcode{28660}
  \country{Spain}
}

\author{N. Juristo}
\email{natalia@fi.upm.es}
\orcid{0000-0002-2465-7141}
\affiliation{%
  \institution{Universidad Polit\'ecnica de Madrid}
  \city{Boadilla del Monte}
  \postcode{28660}
  \country{Spain}
}

\renewcommand{\shortauthors}{O. Dieste, et al.}

\begin{abstract}
\textbf{Background:} Test suites are frequently used to quantify relevant software attributes, such as quality or productivity. \textbf{Problem:} We have detected that the same response variable, measured using different test suites, yields different experiment results. \textbf{Aims:} Assess to which extent differences in test case construction influence measurement accuracy and experimental outcomes. \textbf{Method:} Two industry experiments have been measured using two different test suites, one generated using an \textit{ad-hoc} method and another using \textit{equivalence partitioning}. The accuracy of the measures has been studied using standard procedures, such as ISO 5725, Bland-Altman and Interclass Correlation Coefficients. \textbf{Results:} There are differences in the values of the response variables up to $\pm 60\%$, depending on the test suite (\textit{ad-hoc} vs. \textit{equivalence partitioning}) used. \textbf{Conclusions:} The disclosure of datasets and analysis code is insufficient to ensure the reproducibility of SE experiments. Experimenters should disclose all experimental materials needed to perform independent measurement and re-analysis.
\end{abstract}



%
\maketitle

\input {TEXsections/introduction}
\input {Rsections/problem}

\input {TEXsections/objectives}
\input {Rsections/comparison}

\input {Rsections/results}

\input {Rsections/discussion}
\input {TEXsections/conclusions}

\section{Acknowledgments}

This work was partially supported by the Spanish Ministry of Economy and Competitiveness research grant TIN2014-60490-P, and the Finish TEKES research grant ESEIL (FiDiPro scholarship).

\balance

\bibliographystyle{ACM-Reference-Format}
\bibliography{paper,fernando}


\end{document}

%% file: TEXsections/introduction.tex
\section{Introduction}\label{sec:introduction}

Test-driven development (TDD) research frequently uses the external quality (QLTY) and productivity (PROD) response variables. QLTY is typically measured as the ''amount'' of correct functionality delivered by the developers' code. PROD has a similar definition but is related to a time frame (e.g., the duration of an experimental session). "Functionality" is an abstract concept, not directly observable. In TDD research, test cases are often used as surrogates of functionality.

We have conducted a family of experiments on TDD, as part of the Empirical Software Engineering Industry Lab (ESEIL) project. We used different test suites, as recommended by Shadish et al. \cite[81-82]{Shadish2002}, to measure QLTY and PROD values, thus preventing the mono-method threat to validity. We anticipated some variability among measures, but differences were much larger than we expected. The experimental analyses yield different results, sometimes reversing the effect of the independent variables, depending on the test suite used \cite{Elizabeth2015}.

This paper aims at evaluating to what extent test cases influence the measurement of response variables in TDD experiments. Although the discussion is specifically framed in TDD research, measurement using test cases is frequent in software engineering (SE) research, e.g., \cite{kieburtz1996software,knight1986experimental,feldt1998generating}; other SE areas can thus benefit from our findings. 

The contributions of this paper are:
\begin{itemize}

\item We show that the results of TDD experiments vary depending on the test suites used as measuring instruments. We have assessed this fact in our experiments, but we are certain that the same harmful effect happens in other TDD experiments.

\item We introduce specialized terminology and methods, borrowed from Metrology, the Natural and the Social sciences, to study the accuracy of the test suites when used as measuring instruments.

\item We assess that the measures made using different test suites yield very different results. The same piece of code may exhibit $\pm 60\%$ score differences depending on the test suite used.

\item The publication of datasets and analysis code, as currently required by some publishers, may be sufficient for ensuring reproducibility \cite{NAP25303,fernandez2019open}, but insufficient to evaluate the influence of the measuring instruments. We propose some recommendations to improve the situation: (1) experiments should disclose all experimental materials needed to perform independent measurements, and (2) the practice of re-analysis \cite{mittelstaedt1984econometric,IJzendoorn1994} should be adopted in SE to improve experimental research. 

\end{itemize}

This paper has been written using reproducible research principles. The manuscript \LaTeX~code  is available at \url{https://github.com/GRISE-UPM/TestSuitesMeasurement} (including data files, Java and R code). Analyses have been carried out using R \cite{R} version 4.0.2 (2020-06-22), and the packages \textit{lme4} \cite{lme4}, \textit{xtable} \cite{xtable}, \textit{texreg} \cite{texreg}, \textit{broom} \cite{broom}, \textit{MethComp} \cite{MethComp}, \textit{Hmisc} \cite{Hmisc}, \textit{emmeans} \cite{emmeans}, and \textit{xlsx} \cite{xlsx}.

The paper is structured as follows: Section~\ref{sec:problem-description} describe the research problem. Section~\ref{sec:objectives} sets out the research goals. In Section~\ref{sec:comparison} we introduce the terminology and methods used in Metrology and other sciences for the comparison of measuring instruments. The actual comparison is performed in Section~\ref{sec:comparison-results}. We discuss the implication of our findings in Section~\ref{sec:discussion} and, finally, provide some recommendations in Section~\ref{sec:conclusions}.

%% file: Rsections/problem.tex
\setkeys{Gin}{width=0.60\textwidth}

\section{Problem description}\label{sec:problem-description}

In this paper, we are using two replications conducted in the industry as running examples. These replications will be referred to as \textbf{PT} and \textbf{EC} to maintain companies' anonymity. PT and EC replications have been described in detail in \cite{Tosun2019} and \cite{Dieste2021}, respectively.

\subsection{Experimental replications}\label{sec:design}

PT and EC replications explore two programming strategies: TDD and incremental test-last development (ITLD). TDD requires writing tests before production code, whereas ITLD proceeds inversely. The experimental design is described in Table~\ref{tab:experimental-design}, where the programming strategy is a within-subjects factor. 

The programming strategies have been applied on two greenfield experimental tasks, namely Mars Rover API (MR\footnote{MR and BSK task specifications are included in \url{https://github.com/GRISE-UPM/TestSuitesMeasurement/tree/master/experimental_tasks}.\label{foot:web}}) \cite{mr} and Bowling Score Keeper (BSK\footnote{See footnote \ref{foot:web}.}) \cite{bsk}. MR and BSK are crossed across programming strategies to avoid confounding. This type of design is frequent in SE when participants need to receive specific training, and a few experimental subjects are available. 

The assignment of subjects to groups was performed randomly. 17 and 20 experimental subjects participated in PT and EC, respectively. They were programmers with different degrees of experience, employed in the corresponding companies. PT programmers used Java and jUnit, whereas EC ones used C++ and Boost Test.

\begin{table}[htb]
\small
\centering
\caption{Experimental design}
\label{tab:experimental-design}
\begin{tabular}{llll}
 &  & \multicolumn{2}{c}{\textbf{Treatment}} \\ \cline{3-4} 
 &  & (Session 1) & (Session 2) \\
 &  & ITLD & TDD \\ \cline{2-4} 
\multirow{2}{*}{\textbf{Group}} & Group $MR \rightarrow BSK$ & MR & BSK \\
 & Group $BSK \rightarrow MR$ & BSK & MR \\ \cline{2-4} 
\end{tabular}
\end{table}

\subsection{Response variables and measurement procedure}

We studied external quality (QLTY) and productivity (PROD) response variables. QLTY represents the software quality measured in terms of compliance with the software requirements. Similarly, PROD represents the amount of functionality delivered by programmers. These response variables have been frequently explored in TDD research, e.g., \cite{Causevic2012,Desai2009,Erdogmus2005,Fucci2013}, as well as in our previous research, e.g., \cite{Tosun2016,Dieste2017,Tosun2019}.

QLTY and PROD were measured using specifically designed \textit{unit tests suites}\footnote{The test suites are provided as a single Eclipse workspace containing four projects. They are available as one Eclipse workspace at \url{https://github.com/GRISE-UPM/TestSuitesMeasurement/tree/master/test_suites}.}. One test suite was reused from previous experiments \cite{Erdogmus2005,Tosun2016}. It was generated using an \textit{ad-hoc} (AH) strategy and coded in jUnit. Ad-hoc means that a formal procedure to create the test cases has not been used; the authors of the test suites (MR and BSK) applied their best judgment to derive a set of test cases from the functional requirements.

To avoid the mono-method threat to validity \cite[81-82]{Shadish2002}, we designed new test suites (for MR and BSK) using the \textit{equivalence partitioning} (EP) technique, and coded them in jUnit. We applied equivalence partitioning according to \cite[Chapter 4]{Myers2011}; details can be found in \cite{Elizabeth2015}.  Later, both test suites were ported to Boost Test.

The test suites give a percentage (0\%-100\%) as a result. For instance, in the case of QLTY, this percentage represents the degree to which the code complies with the software requirements: A 0\% value means that the code does not satisfy any requirement; a 100\% value means that the code satisfies all requirements. 

\subsection{Characteristics of the MR and BSK test suites}

The AH and EP test suites are composed of a varying number of test classes/methods/assertions, as indicated in Table \ref{tab:characteristics}. BSK's requirements are well defined, hence the AH and EP test suites exhibit strong similarities: They have the same number of test classes (which are roughly equivalent to functional requirements), and a comparable number of assertions. The EP technique provides a perfect correspondence between test methods and assertions.

MR is defined at a high level and misses a stable specification. Consequently, the AH and EP test suites diverge considerably. However, \textit{divergence} does not imply \textit{measurement differences}. The same code can be measured using different test suites and obtain the same measurement results. For instance, the function 
\lstinline!int sum(int a, int b){ return a + b; }! 
gets a 100\% QLTY with both test suites below:
\begin{lstlisting}[caption={Equivalent tests suites, from the measurement viewpoint},label={lst:equivalent}]
public class Suite1{
	@Test
	public void testOnePlusOneGivesTwo() {
	     assertEquals(2, sum(1, 1)); }
}

public class Suite2{
	@Test
	public void testThreePlusTwoGivesFive() {
	     assertEquals(500, sum(300, 200)); }

	@Test
	public void testThreePlusMinusTwoGivesOne() {
	     assertEquals(300, sum(400, -100)); }
}
\end{lstlisting}

\begin{table}[]
\centering
\small
\caption{Characteristics of the AH and EP test suites}
\label{tab:characteristics}
\begin{tabular}{lllcc}
 &  & \textbf{} & \multicolumn{2}{c}{\textbf{Test suite}} \\ \cline{4-5} 
 &  &  & AH & EP \\ \cline{2-5} 
\multirow{6}{*}{\textbf{Task}} & \multirow{3}{*}{MR} & Test classes & 11 & 9 \\
 &  & Test methods & 52 & 32 \\
 &  & Assertions & 89 & 32 \\ \cline{2-5} 
 & \multirow{3}{*}{BSK} & Test classes & 13 & 13 \\
 &  & Test methods & 48 & 72 \\
 &  & Assertions & 55 & 72 \\ \cline{2-5} 
\end{tabular}
\end{table}

From a testing perspective, the AH and EP test suites are largely equivalent. When we exercise the test suites on correct implementations of MR and BSK, the coverage is almost identical, as shown in Table~\ref{tab:coverage}. Statement coverage is virtually 100\% in all cases. Branch coverage is somewhat smaller but exceeds 90\% (except the AH test suite when applied to the MR task, which has an 88\% branch coverage). 

\begin{table}[]
\small
\centering
\caption{Coverage of the AH and EP test suites}
\label{tab:coverage}
\begin{tabular}{lllrr}
 &  &  & \multicolumn{2}{c}{\textbf{Test suite}} \\ \cline{4-5} 
 &  &  & \multicolumn{1}{c}{AH} & \multicolumn{1}{c}{EP} \\ \cline{2-5} 
\multirow{4}{*}{\textbf{Task}} & \multirow{2}{*}{MR} & Statement coverage & 100\% & 100\% \\
 &  & Branch coverage & 88.1\% & 94.4\% \\ \cline{2-5} 
 & \multirow{2}{*}{BSK} & Statement coverage & 100\% & 96.5\% \\
 &  & Branch coverage & 100\% & 94.4\% \\ \cline{2-5} 
\end{tabular}
\end{table}

Given the coverage values, it is reasonable to assume that both test suites give the same or strongly correlated results. Simple correlation analysis can be used to assess convergent validity \cite[p.67]{Oxford2006}. Table~\ref{tab:correlations} shows the results. All correlations are large ( $r > 0.5$, according to Cohen \cite{Cohen1988}), with the only exception of QLTY at PT ($r = 0.41$, quite close to 0.5), and statistically significant (which is remarkable given the limited sample sizes). At the outset, AH and EP test suites seem to provide similar measures; in the case of EC, to a large extent.

\begin{table}[htb]
\small
\centering
\caption{Correlations between \textit{ad-hoc} and \textit{equivalence partitioning} measures for PT and EC}
\label{tab:correlations}
\begin{tabular}{llrr}

\multicolumn{1}{c}{Experiment} &
\multicolumn{1}{c}{Variable} &
\multicolumn{1}{c}{$r$} &
\multicolumn{1}{c}{$p-value$} \\ \hline
\multirow{2}{*}{PT} & QLTY & 0.41 & 0.02 \\
& PROD & 0.67 & \textless 0.001 \\ \hline
\multirow{2}{*}{EC} & QLTY & 0.72 & \textless 0.001 \\
& PROD & 0.82 & \textless 0.001 \\ \hline
\end{tabular}
\end{table}

\subsection{Problem detection}\label{sec:problem}

PT and EC experiments were analyzed as recommended by Vegas et al. \cite{Vegas2016}, i.e., using a mixed model where \textit{Treatment}, \textit{Task} and \textit{Group} are fixed factors, and \textit{Subject} is a random factor embedded within each \textit{Group}. The analysis model using the \textit{lme4} package \cite{lme4} is:

\begin{equation}
\label{eq:analysis-model}
Y \sim Treatment + Task + Group + ( 1 | Subject )   
\end{equation}

where $Y$ can be QLTY or PROD. We will restrict the discussion to the QLTY response variable, but the comments below match PROD's behavior as well. Tables~\ref{tab:qlty-analysis-pt} and \ref{tab:qlty-analysis-ec} show the analysis results for QLTY at PT and EC, respectively. The numbers between parentheses represent the \textit{standard error} of the \textit{fixed effect} located to its left. The degree of statistical significance is reported using asterisks. The differences between the AH and EP test suites are substantial:

\begin{itemize}

\item The \textit{Task} effect \textbf{reverses depending on the test case definition strategy}. For AH, the effect is negative whereas, for EP, the effect is positive. The changes are dramatic in PT's QLTY (from -24.47 to 20.99 percentage points). Differences are statistically significant both for PT and EC experiments.

\item The \textit{Group} effect is \textbf{positive for AH, and void for EP}. The analysis does not give statistically significant results in this case.

\item Fixed effects are \textbf{larger for AH than EP} regardless of the variable (the \textit{Task} at EC is the exception). Standard deviations are also \textbf{larger for AH than EP} in all cases.

\end{itemize}

There is just one coincidence between the AH and EP measurements:

\begin{itemize}

\item The \textit{Treatment} (ITLD vs. TDD) \textbf{is largely unaffected}. The AH and EP test suites give different values, but the sign and the statistical significance is preserved. 
 
\end{itemize}

\begin{table}[htb]
\small
\centering
\caption{Analysis of the QLTY response variable for PT}\label{tab:qlty-analysis-pt}
\begin{tabular}{l c c}
\hline
 & AH & EP \\
\hline
(Intercept)               & $84.76 \; (10.02)^{***}$ & $28.76 \; (6.94)^{***}$ \\
TreatmentTDD              & $-15.93 \;  (9.44)$      & $-6.30 \; (5.05)$       \\
TaskMR                    & $-24.47 \;  (9.44)^{**}$ & $20.99 \; (5.05)^{***}$ \\
Group$MR \rightarrow BSK$ & $16.78 \; (11.14)$       & $.00 \; (8.77)$         \\
\hline
AIC                       & $310.84$                 & $284.87$                \\
Num. obs.                 & $34$                     & $34$                    \\
Var: Subject (Intercept)  & $148.66$                 & $217.63$                \\
Var: Residual             & $754.33$                 & $215.80$                \\
\hline
\multicolumn{3}{l}{\scriptsize{$^{***}p<0.001$; $^{**}p<0.01$; $^{*}p<0.05$}}
\end{tabular}\end{table}

\begin{table}[htb]
\small
\centering
\caption{Analysis of the QLTY response variable for EC}\label{tab:qlty-analysis-ec}
\begin{tabular}{l c c}
\hline
 & AH & EP \\
\hline
(Intercept)               & $55.56 \; (12.24)^{***}$ & $22.91 \; (7.36)^{**}$ \\
TreatmentTDD              & $14.06 \; (12.22)$       & $6.91 \; (6.87)$       \\
TaskMR                    & $-.10 \; (12.22)$        & $18.90 \; (6.87)^{**}$ \\
Group$MR \rightarrow BSK$ & $7.41 \; (12.27)$        & $-.20 \; (7.82)$       \\
\hline
AIC                       & $388.01$                 & $351.10$               \\
Num. obs.                 & $40$                     & $40$                   \\
Var: Subject (Intercept)  & $6.29$                   & $69.72$                \\
Var: Residual             & $1492.43$                & $472.41$               \\
\hline
\multicolumn{3}{l}{\scriptsize{$^{***}p<0.001$; $^{**}p<0.01$; $^{*}p<0.05$}}
\end{tabular}\end{table}

We expected some disagreement between AH and EP, but not such large discrepancies. 

\begin{framed}

In practice, it implies that the experiment outcomes change depending on the test suites used as a measuring instrument, to the point of obtaining contradictory results.

\end{framed}

%% file: TEXsections/objectives.tex
\section{Research questions and methodology}\label{sec:objectives}

\subsection{Research questions}\label{sec:questions}

Test suites are being used routinely as measuring instruments in TDD experiments, e.g., \cite{Causevic2012,Desai2009,Erdogmus2005,Fucci2013}. TDD experiments are being combined through meta-analysis \cite{rafique2012effects}. We have shown that the experimental results are conditional on the test suites. The same applies, indirectly, to the meta-analyses based on those TDD studies. 

We are concerned about the use of test suites as measuring instruments. We aim to evaluate to which degree similar test suites, e.g., with comparable branch coverage, give different measures. \textbf{A better understanding of the role of test suites for measurement will provide decision criteria for the selection or construction, utilization, and sharing of test suites in SE experiments}.

To the best of our knowledge, \textbf{this problem has not been addressed in the SE literature}. Given its relevance for the TDD community (and from a general perspective to the entire empirical SE), this paper sets out the following research questions:

\vspace{0.8mm}

\textbf{RQ1:} \textit{How} can we assess the accuracy of the measures obtained using test suites?

\vspace{0.8mm}

Measurement is a complex process. Scientists and engineers have developed specific procedures to assess the accuracy of measures, and compare measurement instruments. These procedures can be applied to test suites.

\vspace{0.8mm}

\textbf{RQ2:} \textit{How much} do the AH and EP datasets differ from each other?

\vspace{0.8mm}

The statistical analyses in Section~\ref{sec:problem-description} yield clearly different results. However, such results do not provide an indication of the extent to which the AH and PE datasets differ from each other. Common sense suggests that the differences are large, but we miss a concrete description of \textit{how large} they are.

\subsection{Research method}\label{sec:method}

The research questions posed above imply the comparison of two sets of measurements (AH and EP) generated using different test suites (\textit{ad-hoc} and \textit{equivalence partitioning}). 

The comparison of measurements is not new in SE. Quite a few papers address the comparison of metrics, e.g., \cite{basili1981evaluating,zhang2007performance,zhao1998comparison,jiang2008comparing,di2007comparing}. However, these works do not put the metrics themselves into question, but they typically examine their predictive ability to choose the ''best'' metric for a purpose. Other works, e.g., \cite{meneely2012validating} provide metric validation criteria, but these criteria do not include procedures and methods to compare metrics and decide which ones are more accurate. To conclude, \textbf{we miss theoretical foundations to analyze and compare measurements in SE}.

In turn, different scientific disciplines (e.g., Medicine, Psychology, and Metrology particularly) have dealt with the problem of comparing measurements, giving rise to different comparison approaches. To the best of our knowledge, none of them has been used in SE so far.

\textbf{To answer RQ1}, we provide in Section~\ref{sec:comparison} an abridged description of the different comparison approaches that apply to our research problem.

\textbf{To answer RQ2}, we apply in Section~\ref{sec:comparison-results} all suitable comparison procedures to the AH and EP datasets, with a threefold purpose: (1) quantify how large the differences between measurements are, (2) illustrate how the different comparison approaches can be used in practice, and (3) choose the simplest procedure for routinely use in SE.

%% file: Rsections/comparison.tex
\setkeys{Gin}{width=0.60\textwidth}

\section{Comparing measurements}\label{sec:comparison}

In this section, we will answer \textbf{RQ1: \textit{How} can we assess the accuracy of the measures obtained using test suites?}

Depending on the scientific area, different strategies to compare measurement methods are used. Engineers and Natural Science practitioners are probably acquainted with the approaches advocated by JCGM (Joint Committee for Guides in Metrology) and ISO (International Standards Organization). In the Health Sciences, the Bland-Altman plot and the Intraclass Correlation Coefficient (ICC) are often used. The ICC has also been used in Psychology and the Social Sciences.

\subsection{Fundamental concepts}

Measure theory is the branch of mathematics dealing with the definition and properties of measures \cite{weissteinMeasureTheory}. In SE, ''measures'' are often referred to as ''metrics''. Albeit interchangeable in practice, we will use the term ''measure'' due to its specificity. Fenton and Bieman clarify the differences between both concepts \cite[pp.120-121]{fenton2014software}).

Metrology is the scientific counterpart, specifically interested in the practical implementation of measurements \cite{bipm-metrology} and, more importantly for our purposes, the \textbf{comparison of measurements}. 

Measure theory has been the target of a substantial amount of research in SE. In turn, metrology has been overlooked (with few exceptions such as \cite{abran2002measurement,abran2004metrology}), giving rise to the absence of a methodological background to perform the comparisons of measurements. To fill this gap, we introduce specialized terminology in the following Sections and, later, the comparison methods themselves.

\subsubsection{Basic definitions}\label{sec:basic}

Metrology uses a standard vocabulary, collected in the VIM\footnote{ISO 3534-1 \cite{ISO3534} is an ISO standard that provides definitions similar in most respects to VIM. Other bodies, e.g., national standardization agencies may have defined their vocabularies, typically closely related to VIM.} (Vocabulaire International de M\'etrologie) \cite{VIM}.

According to the VIM, a \textit{measurement} (2.1)\footnote{We include the VIM definition number between parentheses, so the reader can trace back to the standard easily.} is the "process of experimentally obtaining one or more quantity values" from a \textit{measurand} (2.3). 

\begin{framed}
In our case, the measurand is C++ or Java code satisfying some requirements specification (MR, BSK), and the measurement is the code's QLTY.
\end{framed}

Measurement is conducted according to some \textit{measurement method} (2.5), which describes the "logical organization of operations used in a measurement". Among other components, a measurement method includes a \textit{measuring instrument} and a \textit{measurement procedure}. A measuring instrument (3.1) is a "device used for making measurements, alone or in conjunction with one or more supplementary devices".

\begin{framed}
The \textit{ad-hoc} and \textit{equivalence partitioning} test suites are \textbf{measuring instruments} aimed at obtaining QLTY measurements. These instruments were used in conjunction with Eclipse and jUnit/Boost Test frameworks.
\end{framed}

A \textit{measurement procedure} (2.6) is a "detailed description of a measurement", typically intended for a human operator. The measurement method corresponds with the utilization of measurement instruments in practice. 

\begin{framed}

Measurement was performed by one researcher (F. Uyaguari) and involves several steps: (1) connecting the subjects' code with the test suites, resolving syntactic and semantic disagreements, (2) running the test suites, and (3) collecting the pass/failure information for the test cases.

\end{framed}

\subsubsection{Accuracy}\label{sec:accuracy}

\begin{figure}[!t]
\centering
\includegraphics{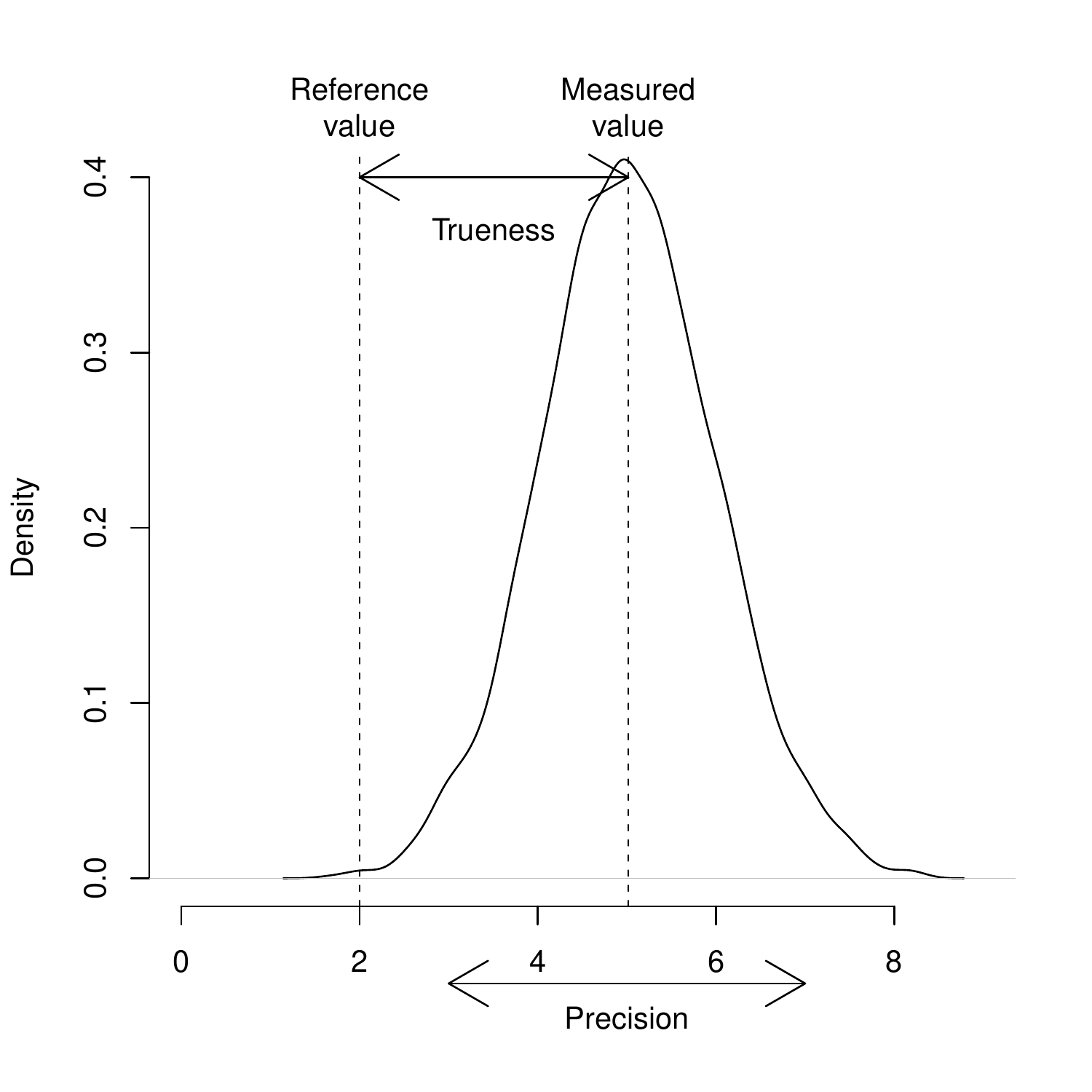}
\caption{Components of measurement \textit{accuracy} according to VIM. This figure assumes that the distribution of the random error is \textit{gaussian}, which is usual but not necessarily true in all circumstances}
\label{fig:accuracy}
\end{figure}

For this research, the concept of \textit{accuracy} is particularly relevant. Accuracy (2.13) is the "closeness of agreement between a measured quantity value and a true quantity value of a measurand". Accuracy has two components: \textit{trueness} and \textit{precision}, , as shown in Fig.~\ref{fig:accuracy}.

\begin{itemize}

\item \textit{Trueness} (2.14) is the "closeness of agreement between the average of an infinite number of replicate measured quantity values and a reference quantity value". Trueness has its origin in the presence of \textit{systematic errors} (2.17), also known as \textit{bias} (2.18).

\item \textit{Precision} (2.15) is the "closeness of agreement between [...] measured quantity values obtained by replicate measurements on the same or similar objects under specified conditions". Precision emerges due to the existence of \textit{random errors} (2.19).

\end{itemize}

\subsection{Determination of accuracy in Engineering and the Natural Sciences}

The determination of accuracy can be performed in different circumstances or \textit{measurement conditions}: (1) Repeatability; (2) Intermediate precision, and (3) Reproducibility.

\subsubsection{Repeatability condition}\label{sec:repeatability}

\textit{Repeatability} (2.20) is the \textit{uncertainty} (2.26) of a set of measurements conducted using "the same measurement procedure, same operators, \textbf{ same measuring [instrument]}, same operating conditions and same location [...] on the same or similar objects over a \textbf{short period of time}". Repeatability is the metrology concept behind the saying ''Measure thrice, cut once'' \cite[p. 3]{Bell2001}, i.e., the concept of repeatability captures the random variability in the measurement process. 

The authoritative source for repeatability analysis is the \textit{Guide to the Expression of Uncertainty in Measurement} (GUM) \cite{GUM}. GUM defines the repeatability uncertainty (denoted $s_r$) as the standard deviation\footnote{In some cases, measurements are not obtained directly. For instance, determining the distance to a distant object using parallax is an indirect measurement, based on two direct measurements: length and angle. The contributions of the length and angle measurements to the uncertainty should be independently determined and later combined using a formula related to Eq.~\ref{eq:propagation-law}; see \cite[Section 5]{GUM} for details.} 
of a set of measures $Y = \{y_1, y_2, ..., y_n\}$:
\begin{equation}
\label{eq:repeatability-expression}
s_r(Y)=\sqrt{\frac{1}{n-1}\sum\limits_{i = 1}^n {(y_i - \bar{y})^2}}
\end{equation}

In this research, $Y$ represents the response variables QLTY and PROD. $\bar{y}$ is the average of all measures, that is, the \textit{measured value} represented in Fig.~\ref{fig:accuracy}. $s_r$ represents the \textit{precision} component of the accuracy. The \textit{trueness} component cannot be captured by repeated measurement because the measurement method can be biased. Bias can be removed by calibration, that is, comparing the measurand with a \textit{reference value} (5.18). For instance, lengths can be traced back to the International Bureau of Weights and Measures' meter bar.
Reference values are unusual but not impossible in SE. For instance, when we measure the correctness of a program against a specification (our research problem) we could code reference programs of known quality, e.g, programs satisfying none or all requirements. Any of these programs could be used as the reference value (at least in principle; see section \ref{sec:discussion} for a brief discussion). 
Trueness would thus be calculated as:
\begin{equation}
\label{eq:bias-expression}
\text{trueness} = \bar{y} - \text{reference value}
\end{equation}

When a reference value is not available, trueness cannot be calculated. However, the repeatability uncertainty is not affected, because the standard deviation is insensitive to location change, i.e., 
\begin{equation}
\label{eq:location-independence}
s_r(Y) = s_r(Y - \text{trueness})
\end{equation}

In practice, repeatability analysis assumes that the instruments are not biased.

\subsubsection{Intermediate precision analysis}\label{sec:intermediate}

The \textit{intermediate precision} (2.22) is a "condition of measurement [...] that includes the same measurement procedure, same location, and replicate measurements on the same or similar objects over an \textbf{extended period of time}, but may include other conditions involving changes". The changes "can include [...] operators, and \textbf{measuring [instruments]}".

The intermediate precision\footnote{The intermediate precision is also termed "within-lab reproducibility" \cite[p. 1]{magnusson2003handbook} because it addresses the variability that happens inside a measuring facility. This stands in contrast to the "between-lab reproducibility", described in Section~\ref{sec:reproducibility}.} deals with two different measurement situations:

\begin{itemize}

\item The variability in the measures that take place over time naturally, e.g., due to varying temperatures throughout the year.

\item The changes in the measurement environment. A typical scenario is the usage of different instruments, e.g., thermometers, to perform the measurements. 

\end{itemize}

\begin{framed}

Our research addresses the second situation. The measuring instruments (the \textit{ad-hoc} and \textit{equivalence partitioning} test suites) have different precisions that influence the overall precision of the measurements.

\end{framed}

The intermediate precision, denoted $s_{R_w}$\cite[p. 1]{magnusson2003handbook}, is calculated as\footnote{The square root is one of the realizations of the \textit{propagation law} defined in the GUM \cite{GUM}. It applies when several ($\geq 2$) independent uncertainties are added.}:
\begin{equation}
\label{eq:propagation-law}
s_{R_w} = \sqrt{s^2_M + s^2_r}
\end{equation}

where $s^2_r$ is the reproducibility uncertainty described in the previous Section and $s^2_M$ is the uncertainty due to the measuring instrument. There are several recommendations to calculate $s^2_M$: the GUM \cite{GUM}, NordTest TR 537 \cite{magnusson2003handbook}, and ISO 5725-3 \cite{ISO5725}. The later is particularly useful for its simplicity. The intermediate precision is calculated using the nested model\footnote{Notice that the same programs are measured using AH and EP}:
\begin{equation}
\label{eq:linear-model-intermediate}
Y = Program/Instrument + \epsilon
\end{equation}
The $Instrument$ is the random factor that represents the measuring instruments implemented with the \textbf{AH and EP test suites}. $s_M$ is given by the associated component of variance, that we will describe in Section \ref{sec:comparison-results}. $\epsilon$ represents the lack of precision that \textit{cannot} be assigned to the $Instrument$, i.e., $\epsilon = s^2_r$ (the repeatability uncertainty).

\subsubsection{Reproducibility uncertainty}\label{sec:reproducibility}

Reproducibility (2.24) is the uncertainty of a set of measurements performed on "\textbf{different condition[s] of measurement}, out of a set of conditions that includes different locations, operators, measuring [instruments], and replicate measurements on the same or similar objects".

The calculation of the reproducibility uncertainty (denoted $s_R$) is described in ISO 5725-2 \cite{ISO5725}. In this part of the standard, reproducibility uncertainty is defined as the uncertainty due to the lab where measurements are conducted. This assumption does not match our research problem, so we will not elaborate it further. However, it could be relevant in other measurement comparison scenarios in SE, e.g., when different research groups participate in the measurement of a multi-site experiment.

\subsubsection{Standard vs. expanded uncertainties}\label{sec:expanded}

$s_r$, $s_{R_w}$ and $s_R$ are standard uncertainties, that is, the $\sigma$ parameter of the normal distribution displayed in Fig.~\ref{fig:accuracy}. However, in a normal distribution, a large percentage of values (around 32\%) are more than 1 standard deviation apart the average (the measured value in Fig.~\ref{fig:accuracy}). 
The \textit{expanded uncertainty} (2.35) provides more significant information about the degree to which measures may differ. Expanded uncertainties are the limits of the interval which includes $(1 - \alpha) \times 100\%$  of the differences among measures. 

In the case of this research (2 measuring instruments, $n$ pieces of code, and one measure per instrument) the expanded uncertainty\footnote{The same formula applies to $s_r$ and $s_R$} would be:
\begin{equation}
\label{eq:expanded-uncertainty1}
t_{(1-\alpha/2), (n-2)} \times s_{R_w}
\end{equation}

$k = t_{(1-\alpha/2), (n-2)}$ is known as \textit{coverage factor} (2.38). When $n \geq 30$, the normal approximation can be used. Typically, $\alpha$ = 0.05, and $k = Z_{(1-\alpha/2)} = 1.96$. In practice, $k$ is rounded up to 2.0 \cite[p. 24]{GUM}. The expanded uncertainty is thus defined as:
\begin{equation}
\label{eq:expanded-uncertainty2}
2 \times s_{R_w}
\end{equation}
and represents the fact that any measure can \textbf{differ up to $\pm 2 \times s_{R_w}$ units} from the average \textit{measured value} (see Fig.~\ref{fig:accuracy}).

\subsection{Determination of accuracy in the Health and Social Sciences}

The correlation coefficient has been the usual procedure to compare measurements in the Health and Social Sciences. This procedure is rather well-known and we have used it already in Section~\ref{sec:problem-description}. 

However, the correlation coefficient exhibits several problems as a comparison procedure. For instance, data with poor agreement\footnote{The term ''accuracy'' is not often used in the Health and Social Sciences. When variables have interval/ratio types, the term \textit{reliability} is frequently used; for nominal/ordinal variables, the most common term is \textit{agreement} \cite{Beckstead2011}. The terms \textit{consistency} and \textit{conformity} can also be found as synonyms of precision and trueness \cite{Muller1994}. Nevertheless, the terms vary depending on the source. For instance, Bland and Altman  \cite{Bland1986} use the term "Agreement" instead of "Reliability" with ratio scales.  We will use the terms defined in the VIM \cite{VIM} and reported in Section~\ref{sec:basic}.} can produce high correlations \cite{Bland1986}; this is exactly what we have observed in Section~\ref{sec:problem-description}. These problems recommended the design of alternative procedures, such as the Bland-Altman method and the usage of the ICC. We will describe both of them in the following sections.

\subsubsection{Bland-Altman method}\label{sec:bland-altman-method}

The Bland-Altman method \cite{Bland1986} is the \textit{de facto} standard for the comparison of measuring instruments in medicine. Contrary to the GUM and ISO, several objects (not the \textit{same} or \textit{similar} object) are involved in the measurement. Each object is measured twice using a different instrument. In our research problem, it implies that we would need two sets of measures: $Y_{AH} = \{y_{AH_1}, y_{AH_2}, \dots, y_{AH_n}\}$ and $Y_{EP} = \{y_{EP_1}, y_{EP_2}, \dots, y_{EP_n}\}$, being $1 \leq i \leq n$ different pieces of code.

The Bland-Altman method starts with the calculation of the difference between\footnote{Procedures for more than two measuring instruments have been proposed in the literature, e.g., \cite{jones2011graphical}.} measurements:
\begin{equation}
\label{eq:bland-altman-difference}
d_i=(y_{AH_i} - y_{EP_i})
\end{equation}

Next, the average of the differences $d_i$ is calculated as:

\begin{equation}
\label{eq:bland-altman-mean-difference}
\bar{d}=\frac{1}{n}\sum\limits_{i = 1}^n {(y_{AH_i} - y_{EP_i})}
\end{equation}

and the standard deviation as:

\begin{equation}
\label{eq:bland-altman-sd-difference}
s_d=\sqrt{\frac{1}{n-1}\sum\limits_{i = 1}^n {(d_i - \bar{d})^2}}
\end{equation}

The Bland-Altman method does not assume that the instruments are unbiased; for this reason, $\bar{d}$ represents the mean difference between the measurements obtained with the AH and EP test suites. If the instruments were unbiased, then $\bar{d} = 0$. $s_d$ is the standard deviation of the difference between measures.

\begin{figure}[!t]
\centering
\includegraphics{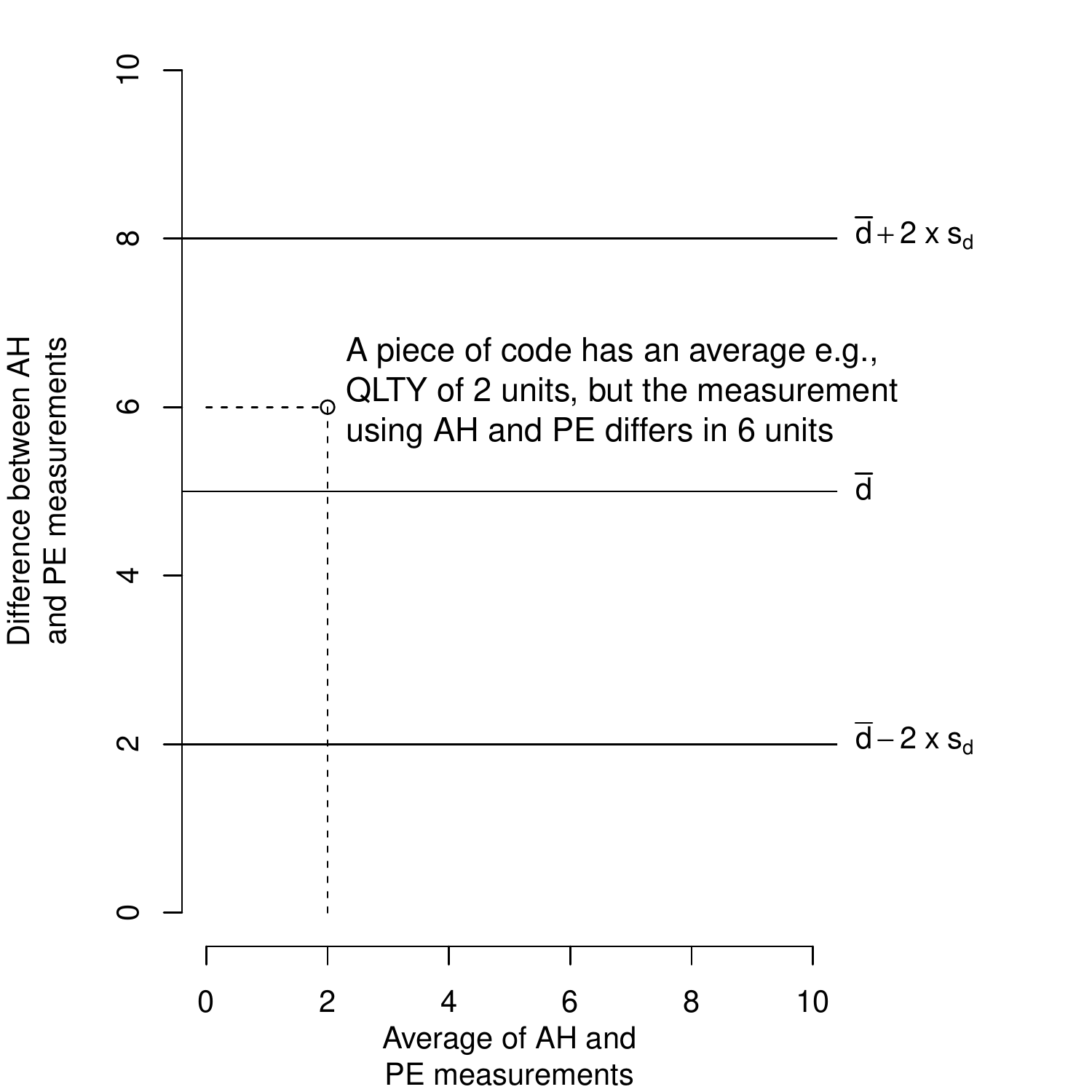}
\caption{Interpretation of the Bland-Altman plot}
\label{fig:example-bland-altman}
\end{figure}

The Bland-Altman method has an associated graphical representation (the Bland-Altman plot), shown in Fig.~\ref{fig:example-bland-altman}. This graph plots the mean values obtained by both measuring instruments:
\begin{equation}
\label{eq:bland-altman-mean}
\frac{y_{AH_i} + y_{EP_i}}{2}
\end{equation}
that is, the best estimation of the true measurement, against their difference:
\begin{equation}
\label{eq:bland-altman-difference}
y_{AH_i} - y_{EP_i}
\end{equation}

A horizontal line is drawn at the mean difference $\bar{d}$. Additionally, the graph also displays two additional horizontal lines located at\footnote{According to Bland and Altman \cite{Bland1986}, either the level $k = 2$ or $k = 1.96$ can be used. We use $k=2$ to highlight the similarities among comparison methods.}:
\begin{equation}
\label{eq:bland-altman-limits}
\bar{d} \pm 2 \times s_d
\end{equation}

Assuming a normal distribution for the differences, these limits enclose 95\% of the differences $d_i$. \textbf{These limits represent how much the measure on the same object varies when one instrument (AH) or another (EP) is used}.

\subsubsection{Intraclass correlation coefficient}\label{sec:icc}

In the Health and Social Sciences, measurements are often scores assigned by human judges. In this case, ''accuracy'' is termed (inter-rater) \textit{agreement}, and represents the degree to which judges coincide with each other when rating the same item. The similarity between ''judges'' and ''measuring instruments'' is apparent. The connection was made for the first time by Lee et al., \cite{lee1989statistical}; they proposed using the ICC as a measure of accuracy.

There are diverse strategies to assess inter-rater agreement\footnote{Refer to \cite{Gwet2014} for an in-depth description.}. For our research problem, the right one is the \textit{Inter-class Correlation Coefficient} (ICC) Model 3 (mixed factorial design), also known as ICC(3, 1) in some statistical packages such as SPSS\textsuperscript{\textregistered} \cite{Gwet2014, Muller1994}. 

The underlying idea is as follows: ICC measures the strength of the relationship among items belonging to some class and compares it to the total variability. In this case, the class is each piece of code. Each class contains two values: the measures taken on that piece of code using both the AH and EP test suites. Mathematically speaking, ICC(3,1) is defined as:
\begin{equation}
\label{eq:icc}
\rho = \frac{s^2_M}{s^2_M + s^2_e}
\end{equation}
where $s^2_M$ is the between-method variance\footnote{The values of $s^2_M$ given by Eqs. \ref{eq:linear-model-intermediate} and \ref{eq:linear-model-icc} are close but not alike. An example appears in section \ref{sec:icc-results}.}, and $s^2_e$ the error variance. $s^2_M$ and $s^2_e$ can be obtained from the linear model:
\begin{equation}
\label{eq:linear-model-icc}
y = Instrument + Program + \epsilon
\end{equation}
whose interpretation is similar to Eq.~\ref{eq:linear-model-intermediate}.

%% file: Rsections/results.tex
\setkeys{Gin}{width=0.60\textwidth}

\section{Comparison of the AH and EP measuring instruments}\label{sec:comparison-results}

In this section, we will answer \textbf{RQ2: \textit{How much} do the AH and EP datasets differ from each other?}

In terms of measurement theory, analyzing the differences between the AH and EP datasets is equivalent to assessing the accuracy (trueness and precision) of the AH and EP measuring instruments. We will use the four comparison methods (repeatability, intermediate precision, Bland-Altman plot and the ICC) described in Section~\ref{sec:comparison}.

\subsection{Repeatability analysis}\label{repeatability-analysis}

Test suites are popular measuring instruments because the measurement is automatic and \textbf{repeatable}. Running the same test suite on the same code yields always the same results\footnote{Varying results are possible when the code depends on some random input. For proper testing in those cases, the random portion should be isolated, e.g., using mocking, to achieve deterministic results \cite{Koskela2013}.}.

The only source of uncertainty when using test suites for measurement is the human operator. To perform the measurement, the operator at least should: (1) download the subject's code, (2) add the AH end EP test suites, (3) make the necessary adjustments to the code (e.g., resolve compilation problems), (4) run the test suites, and (5) write down the results. Measurement problems take place in the steps 3-4. Steps 1-2 influence sample preparation (not measurement), whereas in step 5 only transcription problems may take place.

Step 3 can be performed in different ways. One option is not to make any change to the subjects' code. In this case, due to the repeatable character of the measurement with test suites, the obtained measures have always the same value. This implies that $s_r = 0$. 

\begin{framed}

In the PT and EC experiments, the measurer made small changes (e.g., method names, the order of parameters, etc.) to avoid zero QLTY scores due to clerical errors. Using this strategy, when measurements are repeated \textbf{in a short time}, the results do not vary, because the changes are predictable. Thus, $s_r = 0$ again.

\end{framed}

More complex strategies for connecting the subject's code and the test cases (e.g., fixing loop bounds, order or invocation, etc.) may be more demanding in memory's terms so that the measures do change, giving $s_r > 0$. It does not happen in this research.

\subsection{Intermediate precision}\label{intermediate-precision-analysis}

The intermediate precision $s_{R_w}$ represents the uncertainty produced by the measuring instruments. The intermediate precision assumes that the replicate measurements are made ''on the same or similar object'' (see Section~\ref{sec:intermediate}). This implies a problem in SE experiments. 
We cannot assume that the pieces of code collected in an experiment are similar; in fact, they exhibit a great degree of variation.

Providentially, the nested model proposed in ISO 5725-3 \cite{ISO5725} can be expanded with additional factors, as long as the nesting structure is specified in the model. We have performed the measurement on 74 programs from both the PT and EC experiments\footnote{The origin of the code (PT or EC sites) is irrelevant in the accuracy of the AH and EP test suites}. The Measurement Instrument is nested within the new factor Program. The corresponding model is a simple extension
\footnote{This model can be seen as a restricted version of a more general procedure for the comparison of variances. See \cite[Chapter 9]{box2005statistics} for details.}
of Eq.~\ref{eq:linear-model-intermediate}:
\begin{equation}
\label{eq:linear-model-intermediate-extended}
QLTY = Program/Instrument + \epsilon
\end{equation}

The analysis was conducted with the following R command:

\begin{Schunk}
\begin{Sinput}
> lm <- aov(QLTY ~ Program/Instrument,
+           data = expdata)
\end{Sinput}
\end{Schunk}

\begin{table}[htb]
\small
\centering
\caption{Estimation of $s_M$ using Eq.~\ref{eq:linear-model-intermediate-extended}}
\label{tab:intermediate-precision-analysis}
\begin{tabular}{lrrrrr}
  \hline
 & Df & Sum Sq & Mean Sq & F value & Pr($>$F) \\ 
  \hline
Program & 73.00 & 104051.38 & 1425.36 &  &  \\ 
  Program:Instrument & 74.00 & 71577.93 & 967.27 &  &  \\ 
  Residuals & 0.00 & 0.00 &  &  &  \\ 
   \hline
\end{tabular}\end{table}

Table~\ref{tab:intermediate-precision-analysis} shows the analysis results. The residual is zero, because as we explained above $s_r = 0$. $s^2_M$ is calculated as \cite[pp. 348-350]{box2005statistics}:
\begin{equation}
\label{eq:linear-model-intermediate-extended-component-variance}
s^2_M = MS(Program:Instrument) = 967.27
\end{equation}

The intermediate precision of the measuring instruments is $s_M = \sqrt{s^2_M + s^2_r} = \sqrt{967.27 + 0} = 31.1$. Using $k = 2$, the expanded uncertainty (see Eq.~\ref{eq:expanded-uncertainty2}) is $2 \times 31.1 = 62.2$.

\begin{framed}

When the Ah and EP test suites are used as measuring instruments (e.g., in two different experiments, later combined using meta-analysis), measures that theoretically speaking should be similar (e.g., because the measured programs exhibit the same QLTY) can differ up to $\pm 62.2\%$.

\end{framed}

\subsection{Bland-Altman method}

\begin{figure}[!t]
\includegraphics{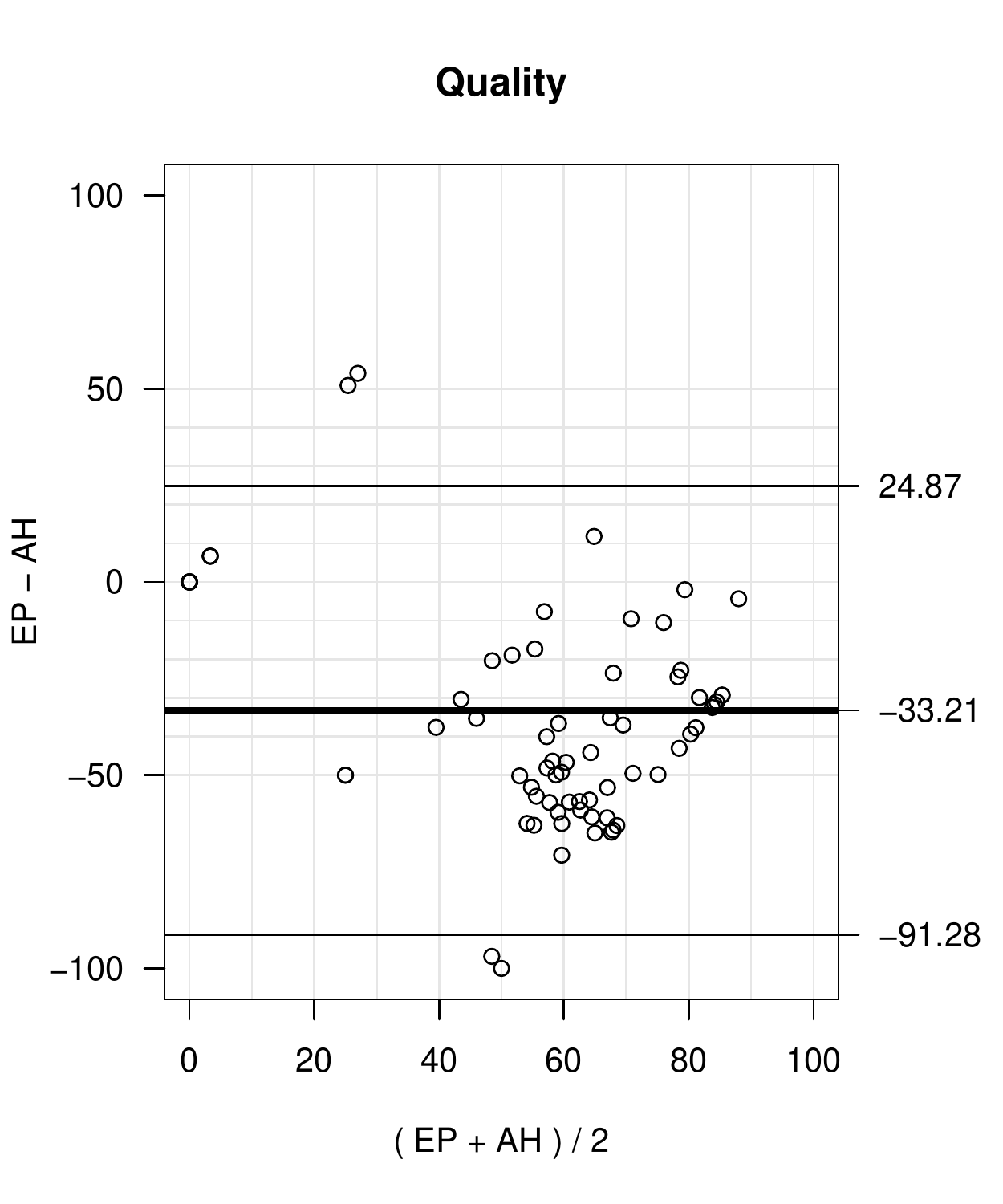}
  \caption{Bland-Altman plot}
  \label{fig:bland-altman-plot}
\end{figure}

The Bland-Altman method uses the differences between measures (Eq.~\ref{eq:bland-altman-difference}) to calculate the accuracy of the measuring instruments. The mean difference $\bar{d} = -33.21$ means that the AH and EP measuring instruments differ 33.21\% units \textit{in average} (notice that QLTY is measured as a percentage). The standard deviation of the differences is $s_d = 29.04$.

Fig.~\ref{fig:bland-altman-plot} shows the same information visually. The central line represents the mean difference $\bar{d} = -33.21$, whereas the top and bottom lines delimit the range of variation of the differences between measurements (the points displayed in the plot). Those limits are calculated as $\bar{d} \pm 2 \times s_d = -33.21 \pm 58.08$.

\begin{framed}

According to the Bland-Altman method, the measures made on the same code by the AH and EP test suites may vary up to 58.08\% in either direction. Actually, we can see in Fig.~\ref{fig:bland-altman-plot} some measures that even exceed such limits. The AH test suite tends to give higher values (33.21\% in average) than the EP test suite.

\end{framed}

\subsection{ICC}\label{sec:icc-results}

The ICC is obtained using Eq.~\ref{eq:icc} which, in turn, requires the calculation of the lineal model depicted in Eq.~\ref{eq:linear-model-icc}. That model is calculated using the R command:

\begin{Schunk}
\begin{Sinput}
> lm <- aov(QLTY ~ 1 +  Instrument + Program,
+           data = expdata)
\end{Sinput}
\end{Schunk}

\begin{table}[htb]
\small
\centering
\caption{Estimation of $s_M$ and $\epsilon$ using Eq.~\ref{eq:linear-model-icc}}
\label{tab:icc-analysis}
\begin{tabular}{lrrrrr}
  \hline
 & Df & Sum Sq & Mean Sq & F value & Pr($>$F) \\ 
  \hline
Instrument & 1.00 & 40803.13 & 40803.13 & 96.79 & 0.00 \\ 
  Program & 73.00 & 104051.38 & 1425.36 & 3.38 & 0.00 \\ 
  Residuals & 73.00 & 30774.79 & 421.57 &  &  \\ 
   \hline
\end{tabular}\end{table}

The results of the analysis is shown in Table~\ref{tab:icc-analysis}. There are strong similarities with Table~\ref{tab:intermediate-precision-analysis} because the models are rather similar. $s_M$ can be calculated from Table~\ref{tab:icc-analysis} as \cite[Table II]{indrayan2013clinical}:
\begin{equation}
\label{eq:icc-component-variance}
s^2_M = \frac{MS(Instrument) - MS(Residual)}{p}
\end{equation}
The notation has the same meaning than in Section~\ref{intermediate-precision-analysis}. $s^2_M$ is thus:
\begin{center}
$s^2_M = \frac{40803.13 - 421.57}{74} = 545.7$
\end{center}

The value of $s^2_M$ is slightly different than the one obtained in Section~\ref{intermediate-precision-analysis} because we are using different linear models in the calculations. The ICC is (see Eq.~\ref{eq:icc}):
\begin{center}
$\rho = \frac{s^2_M}{s^2_M + s^2_e} = \frac{545.7}{545.7 + 421.57} = 0.56$
\end{center}

The interpretation of $\rho$ is not evident. As a general rule, the lower the value, the less related (/similar) the measures taken by the AH and EP test suites for the same program. To interpret the ICC more easily, reference values are typically given in the literature. 

\begin{framed}

Two measurement instruments are considered to have a good agreement when $\rho \geq 0.75$ \cite{Fleiss2011}. In our case, the AH and EP test suites do not achieve that level.

\end{framed}

%% file: Rsections/discussion.tex
\setkeys{Gin}{width=0.60\textwidth}

\section{Discussion}\label{sec:discussion}

\subsection{Accuracy of the AH and EP test suites}

The interpretation of the ICC is contrived. This fact, in addition to some weaknesses \cite{Bland1990,muller1994critical} of the ICC makes this procedure not that useful as a comparison method. In turn, the expanded uncertainties provided by the ISO and Bland-Altman are self-explanatory. 

Regardless of the comparison method used (ISO5725-3, Bland-Altman, or ICC), the accuracy of the AH and EP test suites is rather weak. The Bland-Altman method is particularly illustrative in this regard. The difference between two measures taken on the same program can differ up to 58.08\% in either direction. 

The Bland-Altman plot, shown in Fig.~\ref{fig:bland-altman-plot}, is even more illustrative. Only three points, i.e., three programs, are close to the value ''0'' of the vertical axis, which denotes the coincidence between the measurements obtained with the AH and EP test suites. All other points are 10\%, 20\%, or further apart.

\begin{framed}

The obvious conclusion is that the AH and EP test suites are incompatible measuring instruments. They cannot be used together, because the difference between measurements obtained with each of them is too large. Such differences are likely the origin of the different experimental analysis results that we described in Section~\ref{sec:problem-description}.

\end{framed}

\subsection{Reason of measurement differences}

Measure differences may have multiple origins, some of them minute details. For instance, the test cases in Listing \ref{lst:equivalent} pass for the Java code \lstinline!int sum(int a, int b){ return a + b; }!, but the test case \lstinline!testThreePlusMinusTwoGivesOne()! fails for the C code \lstinline!unsigned char sum(unsigned char a, unsigned char b){ return a + b; }! (in fact, the code probably would not even compile).

The AH and EP test suites are not affected by data types issues, like in the previous example. When the experiments PT and EC were conducted, experimental subjects received code stubs including class and method definitions. The reason for the inconsistent measures lies in the type, and number, of test cases defined in each test suite. 

\begin{figure}[!t]
\centering
\includegraphics{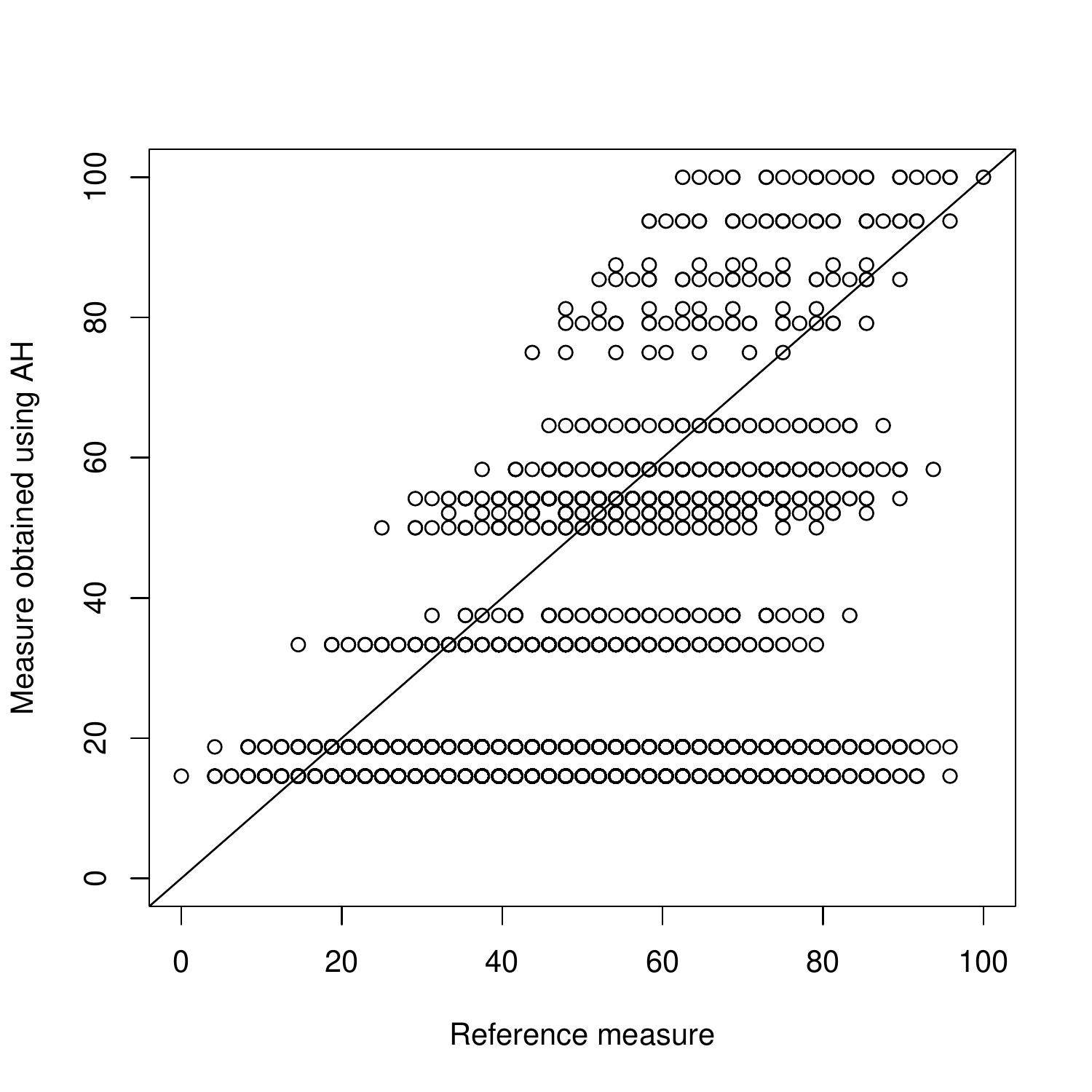}
  \caption{Deviations from reference value (BSK using the AH test suite)}
  \label{fig:deviations_BSK_AH}
\end{figure}

\begin{figure}[!t]
\centering
\includegraphics{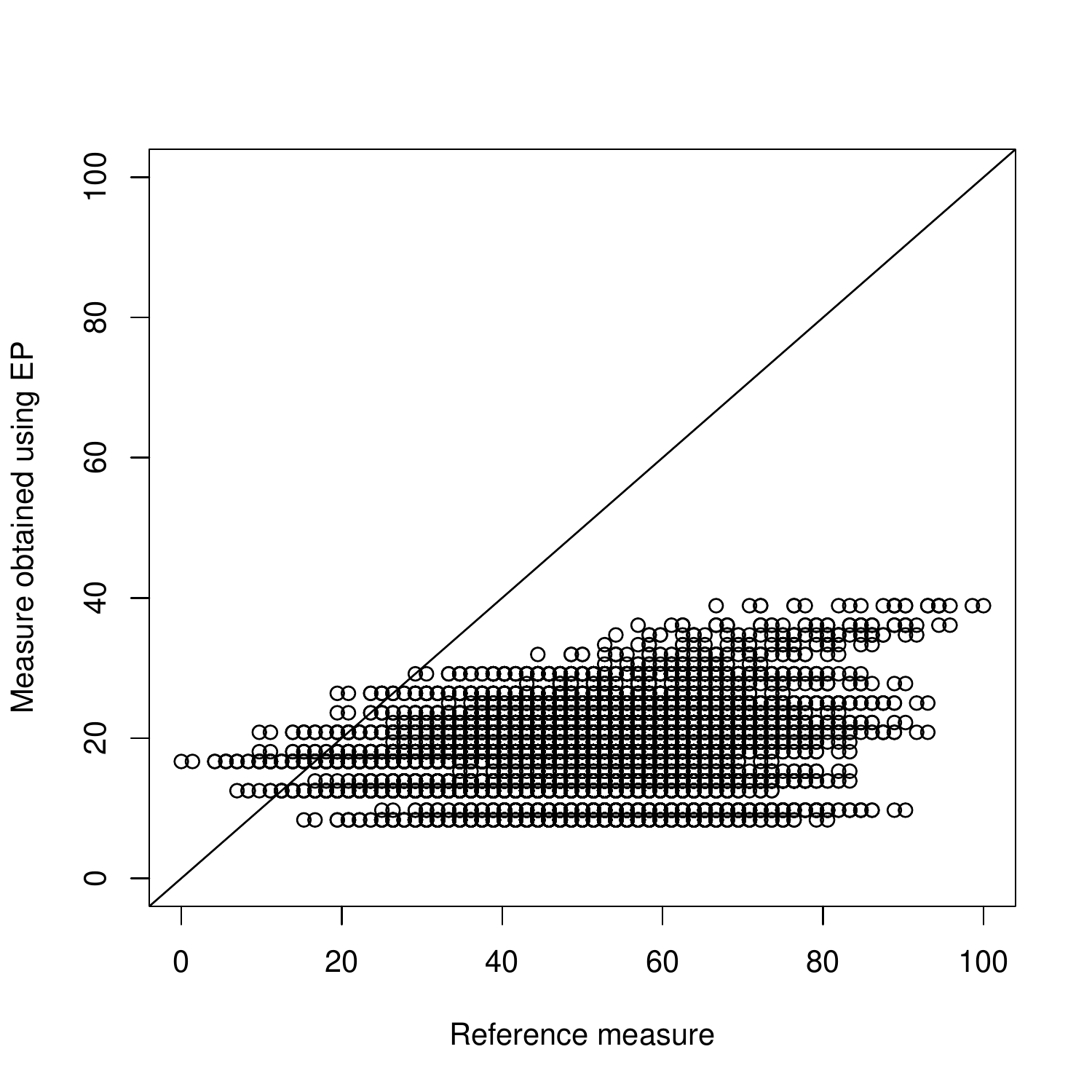}
  \caption{Deviations from reference value (BSK using the EP test suite)}
  \label{fig:deviations_BSK_EP}
\end{figure}

Figure~\ref{fig:deviations_BSK_AH} shows a scatter plot. On the x-axis, we represent the \textit{true value} of a measurement. This true value was obtained using reference code, i.e., code that satisfies all requirements\footnote{The reference code and dataset generation procedures for this section are available as one Eclipse workspace at \url{https://github.com/GRISE-UPM/TestSuitesMeasurement/tree/master/calculation_of_deviations}.}. The task that appears in Figure~\ref{fig:deviations_BSK_AH} is BSK, and the metric displayed is PROD (the plot is easier to understand using PROD instead of QLTY). The y-axis value is the measured value using the AH test suite. If AH provided accurate measures, all points would lie in the diagonal line. Departures from the diagonal line represent measurement errors, the larger the farther apart from the diagonal line the points are. Figure~\ref{fig:deviations_BSK_EP} displays the same information for the EP test suite.

The points in Figure~\ref{fig:deviations_BSK_AH} are scattered around the diagonal line. It implies that AH captures the meaning of the PROD metric. However, individual measures may have large errors. These errors have their primary origin in redundant test cases, i.e., test cases that check the same testing condition. It is fairly easy that \textit{ad-hoc} test case designers insist on multiple testing the same requirement, especially when such requirement is perceived as important. Such test cases pass or fail together, causing large up and down variations in the measured values.

The points in Figure~\ref{fig:deviations_BSK_EP} exhibit a different shape. Most points are located behind the diagonal line, meaning that measured values are systematically lower than true values. Measured PROD values never exceed 40\%. The heuristics of equivalence partitioning testing explain this behavior. Equivalence partitioning puts special emphasis in \textit{invalid classes} which programmers (and \textit{ad-hoc} test case designers) tend to ignore. Actually, the reference code used to create Figures~\ref{fig:deviations_BSK_AH}~and~\ref{fig:deviations_BSK_EP} was obtained from high performing experimental subjects. Overlooking invalid classes lead to systematic low PROD values.

Notice that we are not expressing an opinion about test case design methods. We simply trace (in a simplified manner, as issues may be multiple) the measurement errors to the test suite construction strategies. However, the strategy is not the critical point. The key is that measurement instruments could have been piloted to verify that they produce the right measures, i.e., the ones in the diagonal line, before actual use.

\subsection{Impact in TDD research}

Several synthesis works on TDD have been published recently, e.g., \cite{Rafique2013,Munir,Bissi2016,Causevic2011}. These works identify 90 empirical publications, including surveys, experience reports, case studies, quasi-experiments, and controlled experiments. At least, another 24 studies have been published in the past years, e.g., \cite{Hilton2016,Romano2016}, thus not being included in the synthesis works\footnote{We exclude our own publications, e.g., \cite{Tosun2016} from these figures.}. Out of those 114 publications, 15 experiments measure external quality, i.e., the response variable QLTY used in the PT and EC experiments. 13 out of these 15 experiments use test suites for measurement. The listing is available in Table~\ref{tab:allTDD}. Some patterns are easily noticeable:

\begin{itemize}

\item TDD experiments require subjects to code some experimental task. However, task specifications \textbf{are not usually disclosed}. In some cases, not even the name of the task is reported.

\item Measurement is always performed using test suites. However, with the sole exception of George \& Williams \cite{George2004}, the tests suites \textbf{are not publicly available}.

\item Finally, roughly 50\% of the test suites have been created by the researchers themselves; \textbf{the origin of the remaining 50\% is unknown}. The strategy for test suite creation \textbf{is never reported}.

\end{itemize}

Given the different measures that each test suite yields, TDD experiments may come to different conclusions due to measurement only. The fact that almost 50\% of experiments in Table~\ref{tab:allTDD} use Robert Martin's Bowling Score Keeper (BSK) strengthens our beliefs because our experiments use a modified version of the same task\footnote{See footnote \ref{foot:web}.} and we have already confirmed the impact of the test suite construction strategy on BSK. 

\begin{table*}[htb]
\centering
\caption{Tasks and test suites used by TDD experiments}
\label{tab:allTDD}
\resizebox{!}{0.9\textheight}{%
\begin{adjustbox}{angle=90}
\input{tables/all_tdd_experiments}
\end{adjustbox}
}
\end{table*}

\subsection{Impact in SE research}

Test suites are not used in TDD only. Many experiments, e.g., \cite{kieburtz1996software,knight1986experimental,feldt1998generating}, conducted in other areas of SE, utilize test cases for measurement purposes. It is highly likely that the same dependence between test suites and experimental results take place there too.

%% file: tables/all_tdd_experiments.tex
\begin{tabular}{|p{3cm}|p{1.8cm}|p{7cm}|p{1.8cm}|p{1.8cm}|p{2.5cm}|p{2.5cm}|}
\hline
\textbf{Study} & \textbf{Site} & \textbf{Experimental task} & \textbf{\textbf{Spec avail.?}} & \textbf{Test suite avail.?} & \textbf{Provenance} & \textbf{Construction strategy} \\ \hline
\v Cau\v sevi\'c et al. ~\cite{Causevic2012} & Academy & Robert Martin's Bowling Score Keeper (BSK) & No & No & Not specified & Not specified \\ \hline
Desai et al.~\cite{Desai2009} & Academy & Not specified & No & No & Created by the researcher(s) & Not specified \\ \hline
Erdogmus et al.~\cite{Erdogmus2005} & Academy & Robert Martin's Bowling Score Keeper (BSK) & No & No & Created by the researcher(s) & Not specified \\ \hline
Fucci and Turhan~\cite{Fucci2013} & Academy & Robert Martin's Bowling Score Keeper (BSK) & No & No & Created by the researcher(s) & Not specified \\ \hline
Fucci et al.~\cite{Fucci2016} & Academy & Robert Martin's Bowling Score Keeper (BSK) & No & No & Not specified & Not specified \\ \hline
George and Williams~\cite{George2004} & Industry & Robert Martin's Bowling Score Keeper (BSK) & In B. George's Ph.D. Thesis~\cite{George2002} & In B. George's thesis~\cite{George2002} & Created by the researcher(s) & Not specified \\ \hline
Geras et al.~\cite{Geras2004} & Industry & Program A (registering a new project) + Program B (recording time against a project) & Yes & No & Created by the researcher(s) & Not specified \\ \hline
Gupta and Jalote~\cite{Gupta2007} & Academy & Student registration system / Simple ATM system & Yes & No & Created by the researcher(s) & Not specified \\ \hline
Muller and Hagner~\cite{Muller2002} & Academy & GraphBase (related to graphics) & No & No & Not specified & Not specified \\ \hline
Munir and Moayyed~\cite{Munir2014} & Industry & Robert Martin's Bowling Score Keeper (BSK) & In H. Munir and M. Moayyed's M.Sc. Thesis \cite{Munir2012} & No & Not specified & Not specified \\ \hline
Pan\v cur and al.~\cite{Pancur2003} & Academy & Not specified & No & No & Not specified & Not specified \\ \hline
Pan\v cur and Ciglari\v c~\cite{Pancur2011} & Academy & Distributed database server with built-in data replication mechanism + Chat server & No & No & Created by the researcher(s) & Not specified \\ \hline
Vu et al.~\cite{Vu2009} & Academy & Not specified & No & No & Not specified & Not specified \\ \hline
\end{tabular}

%% file: TEXsections/conclusions.tex
\section{Conclusions}\label{sec:conclusions}

The software metrics area is quite mature in SE. Their formal properties \cite{fenton2014software} and some common pitfalls, e.g., \cite{fenton1992software}, are well understood. However, when dealing with experimental data, it seems that we have overlooked, at least partially, the complexities of measurement. There are several reasons for that: (1) in many cases, the metrics and measuring instruments should be specifically designed for an experiment, (2) the entities of interest in SE are often theoretical constructs, so that objective measurement instruments cannot, by definition, be ever available, (3) we probably trust in excess in the power of statistics, etc.

We have confirmed in this research that different test suites (often used as measuring instruments) give different measures on the same program. Such differences are so radical that they reverse the effects of the factors in the statistical analyses. We have restricted our inquiry to the response variable \textit{external quality}, frequently used in TDD experiments. However, we believe that our findings can be extrapolated to other response variables, research areas and even other research methods, e.g., case studies.

Experimentation is not mature enough to introduce standard measures and measurement instruments. Of course, benchmarks can be adopted. However, the mere adoption of a benchmark does not solve the problems described in this paper because nothing guarantees that such a benchmark provides the \textit{right} measures. Even so, some action should be taken to avoid the harmful effects of metrics and measuring instruments in SE experimental research. In our opinion, three measures can be beneficial for the experimental community:

\begin{itemize}

\item Researchers should disclose not only the measurement results, i.e., the refined data which proceeds to analysis but also the \textbf{raw data} (e.g., subjects' code) and the \textbf{measurement procedure and instruments}. This enables later critical examination and the conduction of a wide range of replications, in particular, re-analysis \cite{mittelstaedt1984econometric,IJzendoorn1994}, where the raw data is independently re-processed before analysis. 

\item The properties of the measures and measuring instruments should be considered before actual measurement takes place. In many cases, measurement standards, e.g., programs satisfying subsets of requirements, can be easily created well before the experiment is conducted, so that formal analysis and empirical studies are possible.

\item When standards are not available, experimenters could use different measures and instruments to avoid threats to construct validity. Coherent results obtained with different instruments would increase the confidence in the experiment results. Just to cite an example, in the PT and EC experiments, the \textit{Treatment} (ITLD vs. TDD) \textbf{was largely unaffected} by the AH and EP test suites. This fact provides us some relief. If the treatments were unaffected, the meta-analyses and other secondary studies based on these data would produce correct results.

\end{itemize}

Finally, in this research, we have only addressed the influence of the measuring instruments in the measurement results. However, the measurement process, as indicated in Section~\ref{sec:comparison}, has several components. They all (in particular, the measurer and the manipulations before applying the measuring instrument) can influence the measurement result too. The assessment of the impact of those elements will be future research.

%% file: paper.bbl

\begin{thebibliography}{80}


\ifx \showCODEN    \undefined \def \showCODEN     #1{\unskip}     \fi
\ifx \showDOI      \undefined \def \showDOI       #1{#1}\fi
\ifx \showISBNx    \undefined \def \showISBNx     #1{\unskip}     \fi
\ifx \showISBNxiii \undefined \def \showISBNxiii  #1{\unskip}     \fi
\ifx \showISSN     \undefined \def \showISSN      #1{\unskip}     \fi
\ifx \showLCCN     \undefined \def \showLCCN      #1{\unskip}     \fi
\ifx \shownote     \undefined \def \shownote      #1{#1}          \fi
\ifx \showarticletitle \undefined \def \showarticletitle #1{#1}   \fi
\ifx \showURL      \undefined \def \showURL       {\relax}        \fi
\providecommand\bibfield[2]{#2}
\providecommand\bibinfo[2]{#2}
\providecommand\natexlab[1]{#1}
\providecommand\showeprint[2][]{arXiv:#2}

\bibitem[\protect\citeauthoryear{??}{bip}{[n.d.]}]%
        {bipm-metrology}
 \bibinfo{year}{[n.d.]}\natexlab{}.
\newblock
\newblock
\urldef\tempurl%
\url{https://www.bipm.org/en/worldwide-metrology/}
\showURL{%
\tempurl}


\bibitem[\protect\citeauthoryear{??}{mr}{2020}]%
        {mr}
 \bibinfo{year}{2020}\natexlab{}.
\newblock \bibinfo{booktitle}{\emph{Mars Rover Kata}}.
\newblock
\urldef\tempurl%
\url{https://kata-log.rocks/mars-rover-kata}
\showURL{%
\tempurl}


\bibitem[\protect\citeauthoryear{Abran and Sellami}{Abran and Sellami}{2002}]%
        {abran2002measurement}
\bibfield{author}{\bibinfo{person}{Alain Abran} {and} \bibinfo{person}{Asma
  Sellami}.} \bibinfo{year}{2002}\natexlab{}.
\newblock \showarticletitle{Measurement and metrology requirements for
  empirical studies in software engineering}. In \bibinfo{booktitle}{\emph{10th
  International Workshop on Software Technology and Engineering Practice}}.
  IEEE, \bibinfo{pages}{185--192}.
\newblock


\bibitem[\protect\citeauthoryear{Abran, Sellami, and Suryn}{Abran
  et~al\mbox{.}}{2004}]%
        {abran2004metrology}
\bibfield{author}{\bibinfo{person}{Alain Abran}, \bibinfo{person}{Asma
  Sellami}, {and} \bibinfo{person}{Witold Suryn}.}
  \bibinfo{year}{2004}\natexlab{}.
\newblock \showarticletitle{Metrology, measurement and metrics in software
  engineering}. In \bibinfo{booktitle}{\emph{Proceedings. 5th International
  Workshop on Enterprise Networking and Computing in Healthcare Industry (IEEE
  Cat. No. 03EX717)}}. IEEE, \bibinfo{pages}{2--11}.
\newblock


\bibitem[\protect\citeauthoryear{Basili and Phillips}{Basili and
  Phillips}{1981}]%
        {basili1981evaluating}
\bibfield{author}{\bibinfo{person}{Victor~R Basili} {and}
  \bibinfo{person}{Tsai-Yun Phillips}.} \bibinfo{year}{1981}\natexlab{}.
\newblock \showarticletitle{Evaluating and comparing software metrics in the
  software engineering laboratory}.
\newblock \bibinfo{journal}{\emph{ACM SIGMETRICS Performance Evaluation
  Review}} \bibinfo{volume}{10}, \bibinfo{number}{1} (\bibinfo{year}{1981}),
  \bibinfo{pages}{95--106}.
\newblock


\bibitem[\protect\citeauthoryear{Bates, M{\"a}chler, Bolker, and Walker}{Bates
  et~al\mbox{.}}{2015}]%
        {lme4}
\bibfield{author}{\bibinfo{person}{Douglas Bates}, \bibinfo{person}{Martin
  M{\"a}chler}, \bibinfo{person}{Ben Bolker}, {and} \bibinfo{person}{Steve
  Walker}.} \bibinfo{year}{2015}\natexlab{}.
\newblock \showarticletitle{Fitting Linear Mixed-Effects Models Using {lme4}}.
\newblock \bibinfo{journal}{\emph{Journal of Statistical Software}}
  \bibinfo{volume}{67}, \bibinfo{number}{1} (\bibinfo{year}{2015}),
  \bibinfo{pages}{1--48}.
\newblock


\bibitem[\protect\citeauthoryear{Beckstead}{Beckstead}{2011}]%
        {Beckstead2011}
\bibfield{author}{\bibinfo{person}{J.W. Beckstead}.}
  \bibinfo{year}{2011}\natexlab{}.
\newblock \showarticletitle{Agreement, reliability, and bias in measurement:
  Commentary on Bland and Altman (1986; 2010)}.
\newblock \bibinfo{journal}{\emph{International Journal of Nursing Studies}}
  \bibinfo{volume}{48}, \bibinfo{number}{1} (\bibinfo{year}{2011}),
  \bibinfo{pages}{134--135}.
\newblock


\bibitem[\protect\citeauthoryear{Bell}{Bell}{2001}]%
        {Bell2001}
\bibfield{author}{\bibinfo{person}{S.A. Bell}.}
  \bibinfo{year}{2001}\natexlab{}.
\newblock \bibinfo{booktitle}{\emph{A Beginner's Guide to Uncertainty of
  Measurement}}.
\newblock
\urldef\tempurl%
\url{http://www.npl.co.uk/publications/a-beginners-guide-to-uncertainty-in-measurement}
\showURL{%
\tempurl}


\bibitem[\protect\citeauthoryear{Bissi, Neto, and Emer}{Bissi
  et~al\mbox{.}}{2016}]%
        {Bissi2016}
\bibfield{author}{\bibinfo{person}{Wilson Bissi}, \bibinfo{person}{Adolfo
  Gustavo Serra~Seca Neto}, {and} \bibinfo{person}{Maria Claudia
  Figueiredo~Pereira Emer}.} \bibinfo{year}{2016}\natexlab{}.
\newblock \showarticletitle{The effects of {Test Driven Development} on
  internal quality, external quality and productivity: A systematic review}.
\newblock \bibinfo{journal}{\emph{Information and Software Technology}}
  \bibinfo{volume}{74} (\bibinfo{year}{2016}), \bibinfo{pages}{45 -- 54}.
\newblock
\showISSN{0950-5849}
\urldef\tempurl%
\url{https://doi.org/10.1016/j.infsof.2016.02.004}
\showDOI{\tempurl}


\bibitem[\protect\citeauthoryear{Bland and Altman}{Bland and Altman}{1986}]%
        {Bland1986}
\bibfield{author}{\bibinfo{person}{M.J. Bland} {and} \bibinfo{person}{D.G.
  Altman}.} \bibinfo{year}{1986}\natexlab{}.
\newblock \showarticletitle{Statistical methods for assessing agreement between
  two methods of clinical measurement}.
\newblock \bibinfo{journal}{\emph{The Lancet}} \bibinfo{volume}{327},
  \bibinfo{number}{8476} (\bibinfo{year}{1986}), \bibinfo{pages}{307--310}.
\newblock


\bibitem[\protect\citeauthoryear{Bland and Altman}{Bland and Altman}{1990}]%
        {Bland1990}
\bibfield{author}{\bibinfo{person}{M.J. Bland} {and} \bibinfo{person}{D.G.
  Altman}.} \bibinfo{year}{1990}\natexlab{}.
\newblock \showarticletitle{A note on the use of the intraclass correlation
  coefficient in the evaluation of agreement between two methods of
  measurement}.
\newblock \bibinfo{journal}{\emph{Computers in Biology and Medicine}}
  \bibinfo{volume}{20}, \bibinfo{number}{5} (\bibinfo{year}{1990}),
  \bibinfo{pages}{337--340}.
\newblock


\bibitem[\protect\citeauthoryear{Box, Hunter, and Hunter}{Box
  et~al\mbox{.}}{2005}]%
        {box2005statistics}
\bibfield{author}{\bibinfo{person}{George~E.P. Box}, \bibinfo{person}{J.~Stuart
  Hunter}, {and} \bibinfo{person}{William~G. Hunter}.}
  \bibinfo{year}{2005}\natexlab{}.
\newblock \bibinfo{booktitle}{\emph{Statistics for Experimenters: Design,
  Discovery and Innovation} (\bibinfo{edition}{second} ed.)}.
\newblock \bibinfo{publisher}{Wiley}.
\newblock
\showISBNx{978-0-471-71813-0}
\urldef\tempurl%
\url{https://books.google.es/books?id=WOQZPwAACAAJ}
\showURL{%
\tempurl}


\bibitem[\protect\citeauthoryear{Carstensen, Gurrin, Ekstrom, and
  Figurski}{Carstensen et~al\mbox{.}}{2015}]%
        {MethComp}
\bibfield{author}{\bibinfo{person}{Bendix Carstensen}, \bibinfo{person}{Lyle
  Gurrin}, \bibinfo{person}{Claus Ekstrom}, {and} \bibinfo{person}{Michal
  Figurski}.} \bibinfo{year}{2015}\natexlab{}.
\newblock \bibinfo{booktitle}{\emph{MethComp: Functions for Analysis of
  Agreement in Method Comparison Studies}}.
\newblock
\urldef\tempurl%
\url{https://CRAN.R-project.org/package=MethComp}
\showURL{%
\tempurl}
\newblock
\shownote{R package version 1.22.2.}


\bibitem[\protect\citeauthoryear{Causevic, Sundmark, and Punnekkat}{Causevic
  et~al\mbox{.}}{2011}]%
        {Causevic2011}
\bibfield{author}{\bibinfo{person}{A. Causevic}, \bibinfo{person}{D. Sundmark},
  {and} \bibinfo{person}{S. Punnekkat}.} \bibinfo{year}{2011}\natexlab{}.
\newblock \showarticletitle{Factors Limiting Industrial Adoption of {Test
  Driven Development}: A Systematic Review}. In \bibinfo{booktitle}{\emph{2011
  Fourth IEEE International Conference on Software Testing, Verification and
  Validation}}. \bibinfo{pages}{337--346}.
\newblock
\showISSN{2159-4848}
\urldef\tempurl%
\url{https://doi.org/10.1109/ICST.2011.19}
\showDOI{\tempurl}


\bibitem[\protect\citeauthoryear{Causevic, Sundmark, and Punnekkat}{Causevic
  et~al\mbox{.}}{2012}]%
        {Causevic2012}
\bibfield{author}{\bibinfo{person}{A. Causevic}, \bibinfo{person}{D. Sundmark},
  {and} \bibinfo{person}{S. Punnekkat}.} \bibinfo{year}{2012}\natexlab{}.
\newblock \showarticletitle{Test case quality in {T}est {D}riven {D}evelopment:
  A study design and a pilot experiment}. In \bibinfo{booktitle}{\emph{16th
  International Conference on Evaluation Assessment in Software Engineering
  (EASE 2012)}}. \bibinfo{pages}{223--227}.
\newblock
\urldef\tempurl%
\url{https://doi.org/10.1049/ic.2012.0029}
\showDOI{\tempurl}


\bibitem[\protect\citeauthoryear{Cohen}{Cohen}{1988}]%
        {Cohen1988}
\bibfield{author}{\bibinfo{person}{J. Cohen}.} \bibinfo{year}{1988}\natexlab{}.
\newblock \bibinfo{booktitle}{\emph{Statistical Power Analysis for the
  Behavioral Sciences}}.
\newblock \bibinfo{publisher}{Routledge}.
\newblock


\bibitem[\protect\citeauthoryear{Dahl}{Dahl}{2016}]%
        {xtable}
\bibfield{author}{\bibinfo{person}{David~B. Dahl}.}
  \bibinfo{year}{2016}\natexlab{}.
\newblock \bibinfo{booktitle}{\emph{xtable: Export Tables to LaTeX or HTML}}.
\newblock
\urldef\tempurl%
\url{https://CRAN.R-project.org/package=xtable}
\showURL{%
\tempurl}
\newblock
\shownote{R package version 1.8-2.}


\bibitem[\protect\citeauthoryear{Desai, Janzen, and Clements}{Desai
  et~al\mbox{.}}{2009}]%
        {Desai2009}
\bibfield{author}{\bibinfo{person}{Chetan Desai}, \bibinfo{person}{David~S.
  Janzen}, {and} \bibinfo{person}{John Clements}.}
  \bibinfo{year}{2009}\natexlab{}.
\newblock \showarticletitle{Implications of Integrating {Test-driven
  Development} into {CS1/CS2} Curricula}. In
  \bibinfo{booktitle}{\emph{Proceedings of the 40th ACM Technical Symposium on
  Computer Science Education}} \emph{(\bibinfo{series}{SIGCSE '09})}.
  \bibinfo{publisher}{ACM}, \bibinfo{pages}{148--152}.
\newblock
\showISBNx{978-1-60558-183-5}
\urldef\tempurl%
\url{https://doi.org/10.1145/1508865.1508921}
\showDOI{\tempurl}


\bibitem[\protect\citeauthoryear{Di~Martino, Ferrucci, Gravino, and
  Mendes}{Di~Martino et~al\mbox{.}}{2007}]%
        {di2007comparing}
\bibfield{author}{\bibinfo{person}{Sergio Di~Martino},
  \bibinfo{person}{Filomena Ferrucci}, \bibinfo{person}{Carmine Gravino}, {and}
  \bibinfo{person}{Emilia Mendes}.} \bibinfo{year}{2007}\natexlab{}.
\newblock \showarticletitle{Comparing size measures for predicting web
  application development effort: A case study}. In
  \bibinfo{booktitle}{\emph{First International Symposium on Empirical Software
  Engineering and Measurement (ESEM 2007)}}. IEEE, \bibinfo{pages}{324--333}.
\newblock


\bibitem[\protect\citeauthoryear{Dieste, Aranda, Uyaguari, Turhan, Tosun,
  Fucci, Oivo, and Juristo}{Dieste et~al\mbox{.}}{2017}]%
        {Dieste2017}
\bibfield{author}{\bibinfo{person}{Oscar Dieste},
  \bibinfo{person}{Alejandrina~M. Aranda}, \bibinfo{person}{Fernando Uyaguari},
  \bibinfo{person}{Burak Turhan}, \bibinfo{person}{Ayse Tosun},
  \bibinfo{person}{Davide Fucci}, \bibinfo{person}{Markku Oivo}, {and}
  \bibinfo{person}{Natalia Juristo}.} \bibinfo{year}{2017}\natexlab{}.
\newblock \showarticletitle{Empirical evaluation of the effects of experience
  on code quality and programmer productivity: an exploratory study}.
\newblock \bibinfo{journal}{\emph{Empirical Software Engineering}}
  (\bibinfo{year}{2017}), \bibinfo{pages}{1--86}.
\newblock


\bibitem[\protect\citeauthoryear{Dieste, Tosun, Vegas, Santos, Uyaguari,
  Kyykk{\"a}, and Juristo}{Dieste et~al\mbox{.}}{2021}]%
        {Dieste2021}
\bibfield{author}{\bibinfo{person}{Oscar Dieste}, \bibinfo{person}{Ayse Tosun},
  \bibinfo{person}{Sira Vegas}, \bibinfo{person}{Adrian Santos},
  \bibinfo{person}{Fernando Uyaguari}, \bibinfo{person}{Jarno Kyykk{\"a}},
  {and} \bibinfo{person}{Natalia Juristo}.} \bibinfo{year}{2021}\natexlab{}.
\newblock \showarticletitle{The role of \textit{slicing} in test-driven
  development}.
\newblock \bibinfo{journal}{\emph{ACM Transactions on Software Engineering and
  Methodology}} (\bibinfo{year}{2021}).
\newblock
\newblock
\shownote{Submitted for review.}


\bibitem[\protect\citeauthoryear{Dragulescu}{Dragulescu}{2014}]%
        {xlsx}
\bibfield{author}{\bibinfo{person}{Adrian~A. Dragulescu}.}
  \bibinfo{year}{2014}\natexlab{}.
\newblock \bibinfo{booktitle}{\emph{xlsx: Read, write, format Excel 2007 and
  Excel 97/2000/XP/2003 files}}.
\newblock
\urldef\tempurl%
\url{https://CRAN.R-project.org/package=xlsx}
\showURL{%
\tempurl}
\newblock
\shownote{R package version 0.5.7.}


\bibitem[\protect\citeauthoryear{Erdogmus, Morisio, and Torchiano}{Erdogmus
  et~al\mbox{.}}{2005}]%
        {Erdogmus2005}
\bibfield{author}{\bibinfo{person}{H. Erdogmus}, \bibinfo{person}{M. Morisio},
  {and} \bibinfo{person}{M. Torchiano}.} \bibinfo{year}{2005}\natexlab{}.
\newblock \showarticletitle{On the effectiveness of the test-first approach to
  programming}.
\newblock \bibinfo{journal}{\emph{IEEE Transactions on Software Engineering}}
  \bibinfo{volume}{31}, \bibinfo{number}{3} (\bibinfo{date}{March}
  \bibinfo{year}{2005}), \bibinfo{pages}{226--237}.
\newblock


\bibitem[\protect\citeauthoryear{Feldt}{Feldt}{1998}]%
        {feldt1998generating}
\bibfield{author}{\bibinfo{person}{Robert Feldt}.}
  \bibinfo{year}{1998}\natexlab{}.
\newblock \showarticletitle{Generating diverse software versions with genetic
  programming: an experimental study}.
\newblock \bibinfo{journal}{\emph{IEE Proceedings-Software}}
  \bibinfo{volume}{145}, \bibinfo{number}{6} (\bibinfo{year}{1998}),
  \bibinfo{pages}{228--236}.
\newblock


\bibitem[\protect\citeauthoryear{Fenton}{Fenton}{1992}]%
        {fenton1992software}
\bibfield{author}{\bibinfo{person}{Norman Fenton}.}
  \bibinfo{year}{1992}\natexlab{}.
\newblock \showarticletitle{When a software measure is not a measure}.
\newblock \bibinfo{journal}{\emph{Software Engineering Journal}}
  \bibinfo{volume}{7}, \bibinfo{number}{5} (\bibinfo{year}{1992}),
  \bibinfo{pages}{357--362}.
\newblock


\bibitem[\protect\citeauthoryear{Fenton and Bieman}{Fenton and Bieman}{2014}]%
        {fenton2014software}
\bibfield{author}{\bibinfo{person}{N. Fenton} {and} \bibinfo{person}{J.
  Bieman}.} \bibinfo{year}{2014}\natexlab{}.
\newblock \bibinfo{booktitle}{\emph{Software Metrics -- A Rigorous and
  Practical Approach}}.
\newblock \bibinfo{publisher}{CRC Press}.
\newblock


\bibitem[\protect\citeauthoryear{Fern{\'a}ndez, Monperrus, Feldt, and
  Zimmermann}{Fern{\'a}ndez et~al\mbox{.}}{2019}]%
        {fernandez2019open}
\bibfield{author}{\bibinfo{person}{Daniel~M{\'e}ndez Fern{\'a}ndez},
  \bibinfo{person}{Martin Monperrus}, \bibinfo{person}{Robert Feldt}, {and}
  \bibinfo{person}{Thomas Zimmermann}.} \bibinfo{year}{2019}\natexlab{}.
\newblock \showarticletitle{The open science initiative of the Empirical
  Software Engineering journal}.
\newblock \bibinfo{journal}{\emph{Empirical Software Engineering}}
  \bibinfo{volume}{24}, \bibinfo{number}{3} (\bibinfo{year}{2019}),
  \bibinfo{pages}{1057--1060}.
\newblock


\bibitem[\protect\citeauthoryear{Fleiss}{Fleiss}{2011}]%
        {Fleiss2011}
\bibfield{author}{\bibinfo{person}{J.L. Fleiss}.}
  \bibinfo{year}{2011}\natexlab{}.
\newblock \bibinfo{booktitle}{\emph{Design and Analysis of Clinical
  Experiments}}.
\newblock \bibinfo{publisher}{Wiley}.
\newblock


\bibitem[\protect\citeauthoryear{Fucci, Scanniello, Romano, Shepperd, Sigweni,
  Uyaguari, Turhan, Juristo, and Oivo}{Fucci et~al\mbox{.}}{2016}]%
        {Fucci2016}
\bibfield{author}{\bibinfo{person}{Davide Fucci}, \bibinfo{person}{Giuseppe
  Scanniello}, \bibinfo{person}{Simone Romano}, \bibinfo{person}{Martin
  Shepperd}, \bibinfo{person}{Boyce Sigweni}, \bibinfo{person}{Fernando
  Uyaguari}, \bibinfo{person}{Burak Turhan}, \bibinfo{person}{Natalia Juristo},
  {and} \bibinfo{person}{Markku Oivo}.} \bibinfo{year}{2016}\natexlab{}.
\newblock \showarticletitle{An External Replication on the Effects of
  {Test-Driven Development} Using a Multi-site Blind Analysis Approach}. In
  \bibinfo{booktitle}{\emph{Proceedings of the 10th ACM/IEEE International
  Symposium on Empirical Software Engineering and Measurement}}
  \emph{(\bibinfo{series}{ESEM '16})}. \bibinfo{publisher}{ACM}, Article
  \bibinfo{articleno}{3}, \bibinfo{numpages}{10}~pages.
\newblock
\showISBNx{978-1-4503-4427-2}
\urldef\tempurl%
\url{https://doi.org/10.1145/2961111.2962592}
\showDOI{\tempurl}


\bibitem[\protect\citeauthoryear{Fucci and Turhan}{Fucci and Turhan}{2013}]%
        {Fucci2013}
\bibfield{author}{\bibinfo{person}{D. Fucci} {and} \bibinfo{person}{B.
  Turhan}.} \bibinfo{year}{2013}\natexlab{}.
\newblock \showarticletitle{A Replicated Experiment on the Effectiveness of
  {Test-First Development}}. In \bibinfo{booktitle}{\emph{2013 ACM / IEEE
  International Symposium on Empirical Software Engineering and Measurement}}
  \emph{(\bibinfo{series}{ESEM '13})}. \bibinfo{publisher}{ACM},
  \bibinfo{pages}{103--112}.
\newblock
\showISSN{1949-3770}
\urldef\tempurl%
\url{https://doi.org/10.1109/ESEM.2013.15}
\showDOI{\tempurl}


\bibitem[\protect\citeauthoryear{George}{George}{2002}]%
        {George2002}
\bibfield{author}{\bibinfo{person}{Boby George}.}
  \bibinfo{year}{2002}\natexlab{}.
\newblock \bibinfo{booktitle}{\emph{Analysis and Quantification of
  {T}est-{D}riven {D}evelopment Approach}}.
\newblock \bibinfo{publisher}{Master's thesis. North Carolina State
  University}.
\newblock


\bibitem[\protect\citeauthoryear{George and Williams}{George and
  Williams}{2004}]%
        {George2004}
\bibfield{author}{\bibinfo{person}{Boby George} {and} \bibinfo{person}{Laurie
  Williams}.} \bibinfo{year}{2004}\natexlab{}.
\newblock \showarticletitle{A structured experiment of {Test-Driven
  Development}}.
\newblock \bibinfo{journal}{\emph{Information and Software Technology}}
  \bibinfo{volume}{46}, \bibinfo{number}{5} (\bibinfo{year}{2004}),
  \bibinfo{pages}{337 -- 342}.
\newblock
\showISSN{0950-5849}
\urldef\tempurl%
\url{https://doi.org/10.1016/j.infsof.2003.09.011}
\showDOI{\tempurl}


\bibitem[\protect\citeauthoryear{Geras, Smith, and Miller}{Geras
  et~al\mbox{.}}{2004}]%
        {Geras2004}
\bibfield{author}{\bibinfo{person}{A. Geras}, \bibinfo{person}{M. Smith}, {and}
  \bibinfo{person}{J. Miller}.} \bibinfo{year}{2004}\natexlab{}.
\newblock \showarticletitle{A prototype empirical evaluation of {Test Driven
  Development}}. In \bibinfo{booktitle}{\emph{Proceedings. 10th International
  Symposium on Software Metrics, 2004.}} \bibinfo{pages}{405--416}.
\newblock
\showISSN{1530-1435}
\urldef\tempurl%
\url{https://doi.org/10.1109/METRIC.2004.1357925}
\showDOI{\tempurl}


\bibitem[\protect\citeauthoryear{Gupta and Jalote}{Gupta and Jalote}{2007}]%
        {Gupta2007}
\bibfield{author}{\bibinfo{person}{A. Gupta} {and} \bibinfo{person}{P.
  Jalote}.} \bibinfo{year}{2007}\natexlab{}.
\newblock \showarticletitle{An Experimental Evaluation of the Effectiveness and
  Efficiency of the {T}est {D}riven {D}evelopment}. In
  \bibinfo{booktitle}{\emph{First International Symposium on Empirical Software
  Engineering and Measurement}} \emph{(\bibinfo{series}{ESEM 2007})}.
  \bibinfo{pages}{285--294}.
\newblock
\showISSN{1949-3770}
\urldef\tempurl%
\url{https://doi.org/10.1109/ESEM.2007.41}
\showDOI{\tempurl}


\bibitem[\protect\citeauthoryear{Gwet}{Gwet}{2014}]%
        {Gwet2014}
\bibfield{author}{\bibinfo{person}{K.L. Gwet}.}
  \bibinfo{year}{2014}\natexlab{}.
\newblock \bibinfo{booktitle}{\emph{Handbook of Inter-Rater Reliability. The
  Definitive Guide to Measuring the Extent of Agreement Among Raters}
  (\bibinfo{edition}{4} ed.)}.
\newblock \bibinfo{publisher}{Advanced Analytics, LLC}.
\newblock


\bibitem[\protect\citeauthoryear{{Harrell Jr}, with contributions~from
  Charles~Dupont, and many others.}{{Harrell Jr} et~al\mbox{.}}{2018}]%
        {Hmisc}
\bibfield{author}{\bibinfo{person}{Frank~E {Harrell Jr}}, \bibinfo{person}{with
  contributions~from Charles~Dupont}, {and} \bibinfo{person}{many others.}}
  \bibinfo{year}{2018}\natexlab{}.
\newblock \bibinfo{booktitle}{\emph{Hmisc: Harrell Miscellaneous}}.
\newblock
\urldef\tempurl%
\url{https://CRAN.R-project.org/package=Hmisc}
\showURL{%
\tempurl}
\newblock
\shownote{R package version 4.1-1.}


\bibitem[\protect\citeauthoryear{Hilton, Nelson, McDonald, McDonald, Metoyer,
  and Dig}{Hilton et~al\mbox{.}}{2016}]%
        {Hilton2016}
\bibfield{author}{\bibinfo{person}{Michael Hilton}, \bibinfo{person}{Nicholas
  Nelson}, \bibinfo{person}{Hugh McDonald}, \bibinfo{person}{Sean McDonald},
  \bibinfo{person}{Ron Metoyer}, {and} \bibinfo{person}{Danny Dig}.}
  \bibinfo{year}{2016}\natexlab{}.
\newblock \showarticletitle{{TDDViz}: Using Software Changes to Understand
  Conformance to {Test Driven Development}}. In \bibinfo{booktitle}{\emph{Agile
  Processes, in Software Engineering, and Extreme Programming: 17th
  International Conference, XP 2016, Edinburgh, UK, May 24-27, 2016,
  Proceedings}}. \bibinfo{publisher}{Springer International Publishing},
  \bibinfo{pages}{53--65}.
\newblock
\showISBNx{978-3-319-33515-5}
\urldef\tempurl%
\url{https://doi.org/10.1007/978-3-319-33515-5_5}
\showDOI{\tempurl}


\bibitem[\protect\citeauthoryear{Indrayan}{Indrayan}{2013}]%
        {indrayan2013clinical}
\bibfield{author}{\bibinfo{person}{Abhaya Indrayan}.}
  \bibinfo{year}{2013}\natexlab{}.
\newblock \showarticletitle{Clinical agreement in quantitative measurements}.
\newblock In \bibinfo{booktitle}{\emph{Methods of Clinical Epidemiology}}.
  \bibinfo{publisher}{Springer}, \bibinfo{pages}{17--27}.
\newblock


\bibitem[\protect\citeauthoryear{{International Standards
  Organization}}{{International Standards Organization}}{1994}]%
        {ISO5725}
\bibfield{author}{\bibinfo{person}{{International Standards Organization}}.}
  \bibinfo{year}{1994}\natexlab{}.
\newblock \bibinfo{booktitle}{\emph{Accuracy (trueness and precision) of
  measurement methods and results -- Parts 1 to 6}}.
\newblock Number ISO 5725.
\newblock


\bibitem[\protect\citeauthoryear{{International Standards
  Organization}}{{International Standards Organization}}{2014}]%
        {ISO3534}
\bibfield{author}{\bibinfo{person}{{International Standards Organization}}.}
  \bibinfo{year}{2014}\natexlab{}.
\newblock \bibinfo{booktitle}{\emph{Statistics-Vocabulary and Symbols -- Part
  1: General statistical terms and terms used in probability}}.
\newblock Number ISO 3534-1.
\newblock


\bibitem[\protect\citeauthoryear{Jiang, Cuki, Menzies, and Bartlow}{Jiang
  et~al\mbox{.}}{2008}]%
        {jiang2008comparing}
\bibfield{author}{\bibinfo{person}{Yue Jiang}, \bibinfo{person}{Bojan Cuki},
  \bibinfo{person}{Tim Menzies}, {and} \bibinfo{person}{Nick Bartlow}.}
  \bibinfo{year}{2008}\natexlab{}.
\newblock \showarticletitle{Comparing design and code metrics for software
  quality prediction}. In \bibinfo{booktitle}{\emph{Proceedings of the 4th
  international workshop on Predictor models in software engineering}}. ACM,
  \bibinfo{pages}{11--18}.
\newblock


\bibitem[\protect\citeauthoryear{{Joint Committee for Guides in
  Metrology}}{{Joint Committee for Guides in Metrology}}{2008}]%
        {GUM}
\bibfield{author}{\bibinfo{person}{{Joint Committee for Guides in Metrology}}.}
  \bibinfo{year}{2008}\natexlab{}.
\newblock \bibinfo{booktitle}{\emph{Evaluation of measurement data -- Guide to
  the expression of uncertainty in measurement}}.
\newblock Number JCGM 100:2008.
\newblock


\bibitem[\protect\citeauthoryear{{Joint Committee for Guides in
  Metrology}}{{Joint Committee for Guides in Metrology}}{2012}]%
        {VIM}
\bibfield{author}{\bibinfo{person}{{Joint Committee for Guides in Metrology}}.}
  \bibinfo{year}{2012}\natexlab{}.
\newblock \bibinfo{booktitle}{\emph{International vocabulary of metrology --
  Basic and general concepts and associated terms}}.
\newblock Number JCGM 200:2012.
\newblock


\bibitem[\protect\citeauthoryear{Jones, Dobson, and O'Brian}{Jones
  et~al\mbox{.}}{2011}]%
        {jones2011graphical}
\bibfield{author}{\bibinfo{person}{Mark Jones}, \bibinfo{person}{Annette
  Dobson}, {and} \bibinfo{person}{Sue O'Brian}.}
  \bibinfo{year}{2011}\natexlab{}.
\newblock \showarticletitle{A graphical method for assessing agreement with the
  mean between multiple observers using continuous measures}.
\newblock \bibinfo{journal}{\emph{International journal of epidemiology}}
  \bibinfo{volume}{40}, \bibinfo{number}{5} (\bibinfo{year}{2011}),
  \bibinfo{pages}{1308--1313}.
\newblock


\bibitem[\protect\citeauthoryear{Kieburtz, McKinney, Bell, Hook, Kotov, Lewis,
  Oliva, Sheard, Smith, and Walton}{Kieburtz et~al\mbox{.}}{1996}]%
        {kieburtz1996software}
\bibfield{author}{\bibinfo{person}{Richard~B Kieburtz}, \bibinfo{person}{Laura
  McKinney}, \bibinfo{person}{Jeffrey~M Bell}, \bibinfo{person}{James Hook},
  \bibinfo{person}{Alex Kotov}, \bibinfo{person}{Jeffrey Lewis},
  \bibinfo{person}{Dino~P Oliva}, \bibinfo{person}{Tim Sheard},
  \bibinfo{person}{Ira Smith}, {and} \bibinfo{person}{Lisa Walton}.}
  \bibinfo{year}{1996}\natexlab{}.
\newblock \showarticletitle{A software engineering experiment in software
  component generation}. In \bibinfo{booktitle}{\emph{Proceedings of the 18th
  international conference on Software engineering}}. IEEE Computer Society,
  \bibinfo{pages}{542--552}.
\newblock


\bibitem[\protect\citeauthoryear{Knight and Leveson}{Knight and
  Leveson}{1986}]%
        {knight1986experimental}
\bibfield{author}{\bibinfo{person}{John~C Knight} {and}
  \bibinfo{person}{Nancy~G Leveson}.} \bibinfo{year}{1986}\natexlab{}.
\newblock \showarticletitle{An experimental evaluation of the assumption of
  independence in multiversion programming}.
\newblock \bibinfo{journal}{\emph{IEEE Transactions on software engineering}}
  \bibinfo{volume}{12}, \bibinfo{number}{1} (\bibinfo{year}{1986}),
  \bibinfo{pages}{96--109}.
\newblock
\urldef\tempurl%
\url{https://doi.org/10.1109/TSE.1986.6312924}
\showDOI{\tempurl}


\bibitem[\protect\citeauthoryear{Koskela}{Koskela}{2013}]%
        {Koskela2013}
\bibfield{author}{\bibinfo{person}{L. Koskela}.}
  \bibinfo{year}{2013}\natexlab{}.
\newblock \bibinfo{booktitle}{\emph{Effective Unit Testing: A Guide for Java
  Developers}}.
\newblock \bibinfo{publisher}{Manning}.
\newblock
\showISBNx{9781935182573}


\bibitem[\protect\citeauthoryear{Lee, Koh, and Ong}{Lee et~al\mbox{.}}{1989}]%
        {lee1989statistical}
\bibfield{author}{\bibinfo{person}{James Lee}, \bibinfo{person}{David Koh},
  {and} \bibinfo{person}{CN Ong}.} \bibinfo{year}{1989}\natexlab{}.
\newblock \showarticletitle{Statistical evaluation of agreement between two
  methods for measuring a quantitative variable}.
\newblock \bibinfo{journal}{\emph{Computers in biology and medicine}}
  \bibinfo{volume}{19}, \bibinfo{number}{1} (\bibinfo{year}{1989}),
  \bibinfo{pages}{61--70}.
\newblock


\bibitem[\protect\citeauthoryear{Leifeld}{Leifeld}{2013}]%
        {texreg}
\bibfield{author}{\bibinfo{person}{Philip Leifeld}.}
  \bibinfo{year}{2013}\natexlab{}.
\newblock \showarticletitle{{texreg}: Conversion of Statistical Model Output in
  {R} to {\LaTeX} and {HTML} Tables}.
\newblock \bibinfo{journal}{\emph{Journal of Statistical Software}}
  \bibinfo{volume}{55}, \bibinfo{number}{8} (\bibinfo{year}{2013}),
  \bibinfo{pages}{1--24}.
\newblock
\urldef\tempurl%
\url{http://www.jstatsoft.org/v55/i08/}
\showURL{%
\tempurl}


\bibitem[\protect\citeauthoryear{Lenth}{Lenth}{2020}]%
        {emmeans}
\bibfield{author}{\bibinfo{person}{Russell Lenth}.}
  \bibinfo{year}{2020}\natexlab{}.
\newblock \bibinfo{booktitle}{\emph{emmeans: Estimated Marginal Means, aka
  Least-Squares Means}}.
\newblock
\urldef\tempurl%
\url{https://CRAN.R-project.org/package=emmeans}
\showURL{%
\tempurl}
\newblock
\shownote{R package version 1.5.0.}


\bibitem[\protect\citeauthoryear{Magnusson}{Magnusson}{2003}]%
        {magnusson2003handbook}
\bibfield{author}{\bibinfo{person}{Bertil Magnusson}.}
  \bibinfo{year}{2003}\natexlab{}.
\newblock \bibinfo{title}{Handbook for calculation of measurement uncertainty
  in environmental laboratories}.
\newblock
\newblock


\bibitem[\protect\citeauthoryear{Martin}{Martin}{2020}]%
        {bsk}
\bibfield{author}{\bibinfo{person}{Robert Martin}.}
  \bibinfo{year}{2020}\natexlab{}.
\newblock \bibinfo{booktitle}{\emph{The Bowling Game Kata}}.
\newblock
\urldef\tempurl%
\url{http://butunclebob.com/ArticleS.UncleBob.TheBowlingGameKata}
\showURL{%
\tempurl}


\bibitem[\protect\citeauthoryear{Meneely, Smith, and Williams}{Meneely
  et~al\mbox{.}}{2012}]%
        {meneely2012validating}
\bibfield{author}{\bibinfo{person}{Andrew Meneely}, \bibinfo{person}{Ben
  Smith}, {and} \bibinfo{person}{Laurie Williams}.}
  \bibinfo{year}{2012}\natexlab{}.
\newblock \showarticletitle{Validating software metrics: A spectrum of
  philosophies}.
\newblock \bibinfo{journal}{\emph{ACM Transactions on Software Engineering and
  Methodology (TOSEM)}} \bibinfo{volume}{21}, \bibinfo{number}{4}
  (\bibinfo{year}{2012}), \bibinfo{pages}{24}.
\newblock


\bibitem[\protect\citeauthoryear{Mittelstaedt and Zorn}{Mittelstaedt and
  Zorn}{1984}]%
        {mittelstaedt1984econometric}
\bibfield{author}{\bibinfo{person}{Robert~A Mittelstaedt} {and}
  \bibinfo{person}{Thomas~S Zorn}.} \bibinfo{year}{1984}\natexlab{}.
\newblock \showarticletitle{Econometric replication: Lessons from the
  experimental sciences}.
\newblock \bibinfo{journal}{\emph{Quarterly Journal of Business and Economics}}
  (\bibinfo{year}{1984}), \bibinfo{pages}{9--15}.
\newblock


\bibitem[\protect\citeauthoryear{Muller and Hagner}{Muller and Hagner}{2002}]%
        {Muller2002}
\bibfield{author}{\bibinfo{person}{M.~M. Muller} {and} \bibinfo{person}{O.
  Hagner}.} \bibinfo{year}{2002}\natexlab{}.
\newblock \showarticletitle{Experiment about {Test-First} programming}.
\newblock \bibinfo{journal}{\emph{IEE Proceedings - Software}}
  \bibinfo{volume}{149}, \bibinfo{number}{5} (\bibinfo{year}{2002}),
  \bibinfo{pages}{131--136}.
\newblock
\showISSN{1462-5970}
\urldef\tempurl%
\url{https://doi.org/10.1049/ip-sen:20020540}
\showDOI{\tempurl}


\bibitem[\protect\citeauthoryear{M\"uller and B\"uttner}{M\"uller and
  B\"uttner}{1994}]%
        {Muller1994}
\bibfield{author}{\bibinfo{person}{R. M\"uller} {and} \bibinfo{person}{P.
  B\"uttner}.} \bibinfo{year}{1994}\natexlab{}.
\newblock \showarticletitle{A Critical Discussion of Intraclass Correlation
  Coefficients}.
\newblock \bibinfo{journal}{\emph{Statistics in Medicine}}
  \bibinfo{volume}{13} (\bibinfo{year}{1994}), \bibinfo{pages}{2465--2476}.
\newblock


\bibitem[\protect\citeauthoryear{M{\"u}ller and B{\"u}ttner}{M{\"u}ller and
  B{\"u}ttner}{1994}]%
        {muller1994critical}
\bibfield{author}{\bibinfo{person}{Reinhold M{\"u}ller} {and}
  \bibinfo{person}{Petra B{\"u}ttner}.} \bibinfo{year}{1994}\natexlab{}.
\newblock \showarticletitle{A critical discussion of intraclass correlation
  coefficients}.
\newblock \bibinfo{journal}{\emph{Statistics in medicine}}
  \bibinfo{volume}{13}, \bibinfo{number}{23-24} (\bibinfo{year}{1994}),
  \bibinfo{pages}{2465--2476}.
\newblock


\bibitem[\protect\citeauthoryear{Munir and Moayyed}{Munir and Moayyed}{2012}]%
        {Munir2012}
\bibfield{author}{\bibinfo{person}{Hussan Munir} {and} \bibinfo{person}{Misagh
  Moayyed}.} \bibinfo{year}{2012}\natexlab{}.
\newblock \emph{\bibinfo{title}{Systematic Literature Review and Controlled
  Pilot Experimental Evaluation of Test Driven Development (TDD) vs. Test-Last
  Development (TLD)}}.
\newblock \bibinfo{thesistype}{Master's\ thesis}. \bibinfo{school}{Blekinge
  Institute of Technology}.
\newblock


\bibitem[\protect\citeauthoryear{Munir, Moayyed, and Petersen}{Munir
  et~al\mbox{.}}{2014a}]%
        {Munir}
\bibfield{author}{\bibinfo{person}{Hussan Munir}, \bibinfo{person}{Misagh
  Moayyed}, {and} \bibinfo{person}{Kai Petersen}.}
  \bibinfo{year}{2014}\natexlab{a}.
\newblock \showarticletitle{Considering Rigor and Relevance when Evaluating
  {Test Driven Development}: A Systematic Review}.
\newblock \bibinfo{journal}{\emph{Information and Software Technology}}
  \bibinfo{volume}{56}, \bibinfo{number}{4} (\bibinfo{year}{2014}),
  \bibinfo{pages}{375--394}.
\newblock
\showISSN{0950-5849}
\urldef\tempurl%
\url{https://doi.org/10.1016/j.infsof.2014.01.002}
\showDOI{\tempurl}


\bibitem[\protect\citeauthoryear{Munir, Wnuk, Petersen, and Moayyed}{Munir
  et~al\mbox{.}}{2014b}]%
        {Munir2014}
\bibfield{author}{\bibinfo{person}{Hussan Munir}, \bibinfo{person}{Krzysztof
  Wnuk}, \bibinfo{person}{Kai Petersen}, {and} \bibinfo{person}{Misagh
  Moayyed}.} \bibinfo{year}{2014}\natexlab{b}.
\newblock \showarticletitle{An Experimental Evaluation of {T}est {D}riven
  {D}evelopment vs. {T}est-last {D}evelopment with Industry Professionals}. In
  \bibinfo{booktitle}{\emph{Proceedings of the 18th International Conference on
  Evaluation and Assessment in Software Engineering}}
  \emph{(\bibinfo{series}{EASE '14})}. \bibinfo{publisher}{ACM}, Article
  \bibinfo{articleno}{50}, \bibinfo{numpages}{10}~pages.
\newblock
\showISBNx{978-1-4503-2476-2}
\urldef\tempurl%
\url{https://doi.org/10.1145/2601248.2601267}
\showDOI{\tempurl}


\bibitem[\protect\citeauthoryear{Myers, Sandler, and Badgett}{Myers
  et~al\mbox{.}}{2011}]%
        {Myers2011}
\bibfield{author}{\bibinfo{person}{G.J. Myers}, \bibinfo{person}{C. Sandler},
  {and} \bibinfo{person}{T. Badgett}.} \bibinfo{year}{2011}\natexlab{}.
\newblock \bibinfo{booktitle}{\emph{The Art of Software Testing}}.
\newblock \bibinfo{publisher}{Wiley}.
\newblock
\showISBNx{9781118133156}


\bibitem[\protect\citeauthoryear{National Academies~of Sciences and
  Medicine}{National Academies~of Sciences and Medicine}{2019}]%
        {NAP25303}
\bibfield{author}{\bibinfo{person}{Engineering National Academies~of Sciences}
  {and} \bibinfo{person}{Medicine}.} \bibinfo{year}{2019}\natexlab{}.
\newblock \bibinfo{booktitle}{\emph{Reproducibility and Replicability in
  Science}}.
\newblock \bibinfo{publisher}{The National Academies Press},
  \bibinfo{address}{Washington, DC}.
\newblock


\bibitem[\protect\citeauthoryear{Ong and M.H.M. Van~Dulmen}{Ong and M.H.M.
  Van~Dulmen}{2006}]%
        {Oxford2006}
\bibfield{author}{\bibinfo{person}{A.D. Ong} {and} \bibinfo{person}{M.H.M.
  M.H.M. Van~Dulmen}.} \bibinfo{year}{2006}\natexlab{}.
\newblock \bibinfo{booktitle}{\emph{Oxford Handbook of Methods in Positive
  Psychology}}.
\newblock \bibinfo{publisher}{Oxford University Press}.
\newblock
\showISBNx{9780199775095}


\bibitem[\protect\citeauthoryear{Pancur and Ciglaric}{Pancur and
  Ciglaric}{2011}]%
        {Pancur2011}
\bibfield{author}{\bibinfo{person}{Matjaz Pancur} {and} \bibinfo{person}{Mojca
  Ciglaric}.} \bibinfo{year}{2011}\natexlab{}.
\newblock \showarticletitle{Impact of {Test-Driven Development} on
  productivity, code and tests: A controlled experiment}.
\newblock \bibinfo{journal}{\emph{Information and Software Technology}}
  \bibinfo{volume}{53}, \bibinfo{number}{6} (\bibinfo{year}{2011}),
  \bibinfo{pages}{557 -- 573}.
\newblock
\showISSN{0950-5849}
\urldef\tempurl%
\url{https://doi.org/10.1016/j.infsof.2011.02.002}
\showDOI{\tempurl}


\bibitem[\protect\citeauthoryear{Pancur, Ciglaric, Trampus, and Vidmar}{Pancur
  et~al\mbox{.}}{2003}]%
        {Pancur2003}
\bibfield{author}{\bibinfo{person}{M. Pancur}, \bibinfo{person}{M. Ciglaric},
  \bibinfo{person}{M. Trampus}, {and} \bibinfo{person}{T. Vidmar}.}
  \bibinfo{year}{2003}\natexlab{}.
\newblock \showarticletitle{Towards empirical evaluation of {Test-Driven
  Development} in a university environment}. In \bibinfo{booktitle}{\emph{The
  IEEE Region 8 EUROCON 2003. Computer as a Tool}}, Vol.~\bibinfo{volume}{2}.
  \bibinfo{pages}{83--86 vol.2}.
\newblock
\urldef\tempurl%
\url{https://doi.org/10.1109/EURCON.2003.1248153}
\showDOI{\tempurl}


\bibitem[\protect\citeauthoryear{{R Core Team}}{{R Core Team}}{2016}]%
        {R}
\bibfield{author}{\bibinfo{person}{{R Core Team}}.}
  \bibinfo{year}{2016}\natexlab{}.
\newblock \bibinfo{booktitle}{\emph{R: A Language and Environment for
  Statistical Computing}}.
\newblock R Foundation for Statistical Computing, Vienna, Austria.
\newblock
\urldef\tempurl%
\url{https://www.R-project.org/}
\showURL{%
\tempurl}


\bibitem[\protect\citeauthoryear{Rafique and Mi{\v{s}}i{\'c}}{Rafique and
  Mi{\v{s}}i{\'c}}{2012}]%
        {rafique2012effects}
\bibfield{author}{\bibinfo{person}{Yahya Rafique} {and}
  \bibinfo{person}{Vojislav~B Mi{\v{s}}i{\'c}}.}
  \bibinfo{year}{2012}\natexlab{}.
\newblock \showarticletitle{The effects of test-driven development on external
  quality and productivity: A meta-analysis}.
\newblock \bibinfo{journal}{\emph{IEEE Transactions on Software Engineering}}
  \bibinfo{volume}{39}, \bibinfo{number}{6} (\bibinfo{year}{2012}),
  \bibinfo{pages}{835--856}.
\newblock


\bibitem[\protect\citeauthoryear{Rafique and Misic}{Rafique and Misic}{2013}]%
        {Rafique2013}
\bibfield{author}{\bibinfo{person}{Y. Rafique} {and} \bibinfo{person}{V.~B.
  Misic}.} \bibinfo{year}{2013}\natexlab{}.
\newblock \showarticletitle{The Effects of {Test-Driven Development} on
  External Quality and Productivity: A Meta-Analysis}.
\newblock \bibinfo{journal}{\emph{IEEE Transactions on Software Engineering}}
  \bibinfo{volume}{39}, \bibinfo{number}{6} (\bibinfo{year}{2013}),
  \bibinfo{pages}{835--856}.
\newblock
\showISSN{0098-5589}
\urldef\tempurl%
\url{https://doi.org/10.1109/TSE.2012.28}
\showDOI{\tempurl}


\bibitem[\protect\citeauthoryear{Robinson}{Robinson}{2018}]%
        {broom}
\bibfield{author}{\bibinfo{person}{David Robinson}.}
  \bibinfo{year}{2018}\natexlab{}.
\newblock \bibinfo{booktitle}{\emph{broom: Convert Statistical Analysis Objects
  into Tidy Data Frames}}.
\newblock
\urldef\tempurl%
\url{https://CRAN.R-project.org/package=broom}
\showURL{%
\tempurl}
\newblock
\shownote{R package version 0.4.4.}


\bibitem[\protect\citeauthoryear{Romano, Fucci, Scanniello, Turhan, and
  Juristo}{Romano et~al\mbox{.}}{2016}]%
        {Romano2016}
\bibfield{author}{\bibinfo{person}{Simone Romano}, \bibinfo{person}{Davide
  Fucci}, \bibinfo{person}{Giuseppe Scanniello}, \bibinfo{person}{Burak
  Turhan}, {and} \bibinfo{person}{Natalia Juristo}.}
  \bibinfo{year}{2016}\natexlab{}.
\newblock \showarticletitle{Results from an Ethnographically-informed Study in
  the Context of {Test Driven Development}}. In
  \bibinfo{booktitle}{\emph{Proceedings of the 20th International Conference on
  Evaluation and Assessment in Software Engineering}}
  \emph{(\bibinfo{series}{EASE '16})}. \bibinfo{publisher}{ACM}, Article
  \bibinfo{articleno}{10}, \bibinfo{numpages}{10}~pages.
\newblock
\showISBNx{978-1-4503-3691-8}
\urldef\tempurl%
\url{https://doi.org/10.1145/2915970.2915996}
\showDOI{\tempurl}


\bibitem[\protect\citeauthoryear{Sequeda}{Sequeda}{2015}]%
        {Elizabeth2015}
\bibfield{author}{\bibinfo{person}{Dora Elizabeth~Jaimes Sequeda}.}
  \bibinfo{year}{2015}\natexlab{}.
\newblock \bibinfo{title}{{?`}Se producen diferencias considerables en los
  resultados experimentales en funci{\'o}n del procedimiento de medici{\'o}n?
  Un estudio comparativo para los constructos calidad y productividad en el
  marco de un experimento de test-driven development.}  (\bibinfo{date}{Abril}
  \bibinfo{year}{2015}).
\newblock
\urldef\tempurl%
\url{http://oa.upm.es/44268/}
\showURL{%
\tempurl}


\bibitem[\protect\citeauthoryear{Shadish, Cook, and Campbell}{Shadish
  et~al\mbox{.}}{2002}]%
        {Shadish2002}
\bibfield{author}{\bibinfo{person}{W.R. Shadish}, \bibinfo{person}{T.D. Cook},
  {and} \bibinfo{person}{D.T. Campbell}.} \bibinfo{year}{2002}\natexlab{}.
\newblock \bibinfo{booktitle}{\emph{Experimental and Quasi-Experimental Designs
  for Generalized Causal Inference}}.
\newblock \bibinfo{publisher}{Houghton Mifflin Company}. 510 pages.
\newblock


\bibitem[\protect\citeauthoryear{Tosun, Dieste, Fucci, Vegas, Turhan, Erdogmus,
  Santos, Oivo, Toro, Jarvinen, and Juristo}{Tosun et~al\mbox{.}}{2016}]%
        {Tosun2016}
\bibfield{author}{\bibinfo{person}{Ayse Tosun}, \bibinfo{person}{Oscar Dieste},
  \bibinfo{person}{Davide Fucci}, \bibinfo{person}{Sira Vegas},
  \bibinfo{person}{Burak Turhan}, \bibinfo{person}{Hakan Erdogmus},
  \bibinfo{person}{Adrian Santos}, \bibinfo{person}{Markku Oivo},
  \bibinfo{person}{Kimmo Toro}, \bibinfo{person}{Janne Jarvinen}, {and}
  \bibinfo{person}{Natalia Juristo}.} \bibinfo{year}{2016}\natexlab{}.
\newblock \showarticletitle{An industry experiment on the effects of
  test-driven development on external quality and productivity}.
\newblock \bibinfo{journal}{\emph{Empirical Software Engineering}}
  (\bibinfo{year}{2016}), \bibinfo{pages}{1--43}.
\newblock
\showISSN{1573-7616}


\bibitem[\protect\citeauthoryear{{Tosun}, {Dieste}, {Vegas}, {Pfahl}, {Rungi},
  and {Juristo}}{{Tosun} et~al\mbox{.}}{2019}]%
        {Tosun2019}
\bibfield{author}{\bibinfo{person}{A. {Tosun}}, \bibinfo{person}{O. {Dieste}},
  \bibinfo{person}{S. {Vegas}}, \bibinfo{person}{D. {Pfahl}},
  \bibinfo{person}{K. {Rungi}}, {and} \bibinfo{person}{N. {Juristo}}.}
  \bibinfo{year}{2019}\natexlab{}.
\newblock \showarticletitle{Investigating the Impact of Development Task on
  External Quality in Test-Driven Development: An Industry Experiment}.
\newblock \bibinfo{journal}{\emph{IEEE Transactions on Software Engineering}}
  (\bibinfo{year}{2019}).
\newblock
\urldef\tempurl%
\url{https://doi.org/10.1109/TSE.2019.2949811}
\showDOI{\tempurl}


\bibitem[\protect\citeauthoryear{van IJzendoorn}{van IJzendoorn}{1994}]%
        {IJzendoorn1994}
\bibfield{author}{\bibinfo{person}{Marinus~H. van IJzendoorn}.}
  \bibinfo{year}{1994}\natexlab{}.
\newblock \showarticletitle{A process model of replication studies: On the
  relation between different types of replication}.
\newblock In \bibinfo{booktitle}{\emph{Reconstructing the Mind: Replicability
  in research on human development}},
  \bibfield{editor}{\bibinfo{person}{R.~van~der Veer}, \bibinfo{person}{H.H.
  van IJzendoorn}, {and} \bibinfo{person}{J.~Valsiner}} (Eds.).
  \bibinfo{publisher}{Ablex Publishing Corporation}, \bibinfo{address}{New
  Jersey}.
\newblock


\bibitem[\protect\citeauthoryear{Vegas, Apa, and Juristo}{Vegas
  et~al\mbox{.}}{2016}]%
        {Vegas2016}
\bibfield{author}{\bibinfo{person}{S. Vegas}, \bibinfo{person}{C. Apa}, {and}
  \bibinfo{person}{N. Juristo}.} \bibinfo{year}{2016}\natexlab{}.
\newblock \showarticletitle{Crossover Designs in Software Engineering
  Experiments: Benefits and Perils}.
\newblock \bibinfo{journal}{\emph{Transactions on Software Engineering}}
  \bibinfo{volume}{42}, \bibinfo{number}{2} (\bibinfo{year}{2016}),
  \bibinfo{pages}{120--135}.
\newblock


\bibitem[\protect\citeauthoryear{Vu, Frojd, Shenkel-Therolf, and Janzen}{Vu
  et~al\mbox{.}}{2009}]%
        {Vu2009}
\bibfield{author}{\bibinfo{person}{J.~H. Vu}, \bibinfo{person}{N. Frojd},
  \bibinfo{person}{C. Shenkel-Therolf}, {and} \bibinfo{person}{D.~S. Janzen}.}
  \bibinfo{year}{2009}\natexlab{}.
\newblock \showarticletitle{Evaluating {Test-Driven Development} in an
  Industry-Sponsored Capstone Project}. In \bibinfo{booktitle}{\emph{2009 Sixth
  International Conference on Information Technology: New Generations}}.
  \bibinfo{pages}{229--234}.
\newblock
\urldef\tempurl%
\url{https://doi.org/10.1109/ITNG.2009.11}
\showDOI{\tempurl}


\bibitem[\protect\citeauthoryear{Weisstein}{Weisstein}{[n.d.]}]%
        {weissteinMeasureTheory}
\bibfield{author}{\bibinfo{person}{Eric~W. Weisstein}.}
  \bibinfo{year}{[n.d.]}\natexlab{}.
\newblock \bibinfo{booktitle}{\emph{Measure Theory}}.
\newblock
\urldef\tempurl%
\url{http://mathworld.wolfram.com/MeasureTheory.html}
\showURL{%
\tempurl}
\newblock
\shownote{Last visited on 22/5/2019.}


\bibitem[\protect\citeauthoryear{Zhang and Baddoo}{Zhang and Baddoo}{2007}]%
        {zhang2007performance}
\bibfield{author}{\bibinfo{person}{Min Zhang} {and} \bibinfo{person}{Nathan
  Baddoo}.} \bibinfo{year}{2007}\natexlab{}.
\newblock \showarticletitle{Performance comparison of software complexity
  metrics in an open source project}. In \bibinfo{booktitle}{\emph{European
  Conference on Software Process Improvement}}. Springer,
  \bibinfo{pages}{160--174}.
\newblock


\bibitem[\protect\citeauthoryear{Zhao, Wohlin, Ohlsson, and Xie}{Zhao
  et~al\mbox{.}}{1998}]%
        {zhao1998comparison}
\bibfield{author}{\bibinfo{person}{Ming Zhao}, \bibinfo{person}{Claes Wohlin},
  \bibinfo{person}{Niclas Ohlsson}, {and} \bibinfo{person}{Min Xie}.}
  \bibinfo{year}{1998}\natexlab{}.
\newblock \showarticletitle{A comparison between software design and code
  metrics for the prediction of software fault content}.
\newblock \bibinfo{journal}{\emph{Information and Software Technology}}
  \bibinfo{volume}{40}, \bibinfo{number}{14} (\bibinfo{year}{1998}),
  \bibinfo{pages}{801--809}.
\newblock


\end{thebibliography}
